\documentclass[10pt, conference, compsocconf]{IEEEtran}
\usepackage {graphicx}
\usepackage{multirow}
\usepackage{rotating}
\usepackage{mathtools, cuted}
\usepackage{lipsum, color}
\usepackage{booktabs}
\usepackage{array,dcolumn}
\usepackage[flushleft]{threeparttable}
\usepackage[bookmarks=false]{hyperref}
\begin{document}
\title{On the Prediction of $>$100 MeV Solar Energetic Particle Events Using
GOES Satellite Data}

\author{
    \IEEEauthorblockN{ Soukaina Filali Boubrahimi \IEEEauthorrefmark{1},
    Berkay Aydin \IEEEauthorrefmark{1}, Petrus Martens
    \IEEEauthorrefmark{2}, and Rafal Angryk
    \IEEEauthorrefmark{1}}
     \IEEEauthorblockA{\IEEEauthorrefmark{1}Georgia State University, Departement of Computer Science }
     Email: \{sfilaliboubrahimi1, baydin2, rangryk\}@cs.gsu.edu
    \IEEEauthorblockA{\IEEEauthorrefmark{2}Georgia State University,
    Departement of Physics and Astronomy}Email:
    \{martens\}@astro.gsu.edu
     }



\maketitle

\begin{abstract}

Solar energetic particles are a result of intense solar events such as
solar flares and Coronal Mass Ejections (CMEs). These latter events all together
can cause major disruptions to spacecraft that are in Earth's orbit and outside
of the magnetosphere. In this work we are interested in establishing the necessary
conditions for a major geo-effective solar particle storm immediately after a
major flare, namely the existence of a direct magnetic connection. To our
knowledge, this is the first work that explores not only the correlations of
GOES X-ray and proton channels, but also the correlations that happen across all
the proton channels. We found that proton channels
auto-correlations and cross-correlations may also be precursors to the
occurrence of an SEP event. In this paper, we tackle the problem of predicting
$>$100 MeV SEP events from a multivariate time series perspective using easily
interpretable decision tree models.

\end{abstract}

\begin{IEEEkeywords}
SEP Events Prediction; CART decision tree, $>$100 MeV SEP; GOES X-ray and Proton
correlation; Vector autoregression

\end{IEEEkeywords}

\IEEEpeerreviewmaketitle

\section{Introduction}
The occurence of important Solar Energetic Particle (SEP) events is one of the
prominent planning considerations for manned and unmanned lunar and planetary
missions \cite{posner2007up}.
A high exposure to large solar particles events can deliver critical doses to
human organs and may damage the instruments on board of satellites and the
global positioning system (GPS) due to the risk of saturation. SEP
events usually happen 30 minutes after the occurrence of the X-ray flare, which
leaves very little time for astronauts performing extra-vehicular activity on
the International Space Station or planetary surfaces to
take evasive actions \cite{2016l}. Earlier warning of the SEP events will be a valuable tool
for mission controllers that need to take a prompt decision concerning the
atrounauts' safety and the mission completion.
When a solar flare or a CME happens, the magnetic force that is exercised is
manifested through different effects. Some of the effects are listed in their
order of occurrence: light, thermal, particle acceleration, and matter ejection
in case of CMEs. The first effect of a solar flare is a flash of increased
brightness that is
observed near the Sun's surface which is due to the X-rays and UV radiation.
Then, part of the magnetic energy is converted into thermal energy in the area
where the eruption happened.
Solar particles in the atmosphere are then accelerated with different speed
spectra, that can reach up to 80\% of the speed of light, depending on the
intensity of the parent eruptive event.
  \begin{figure}[h!]
    \centering
    \includegraphics[width=1\linewidth]{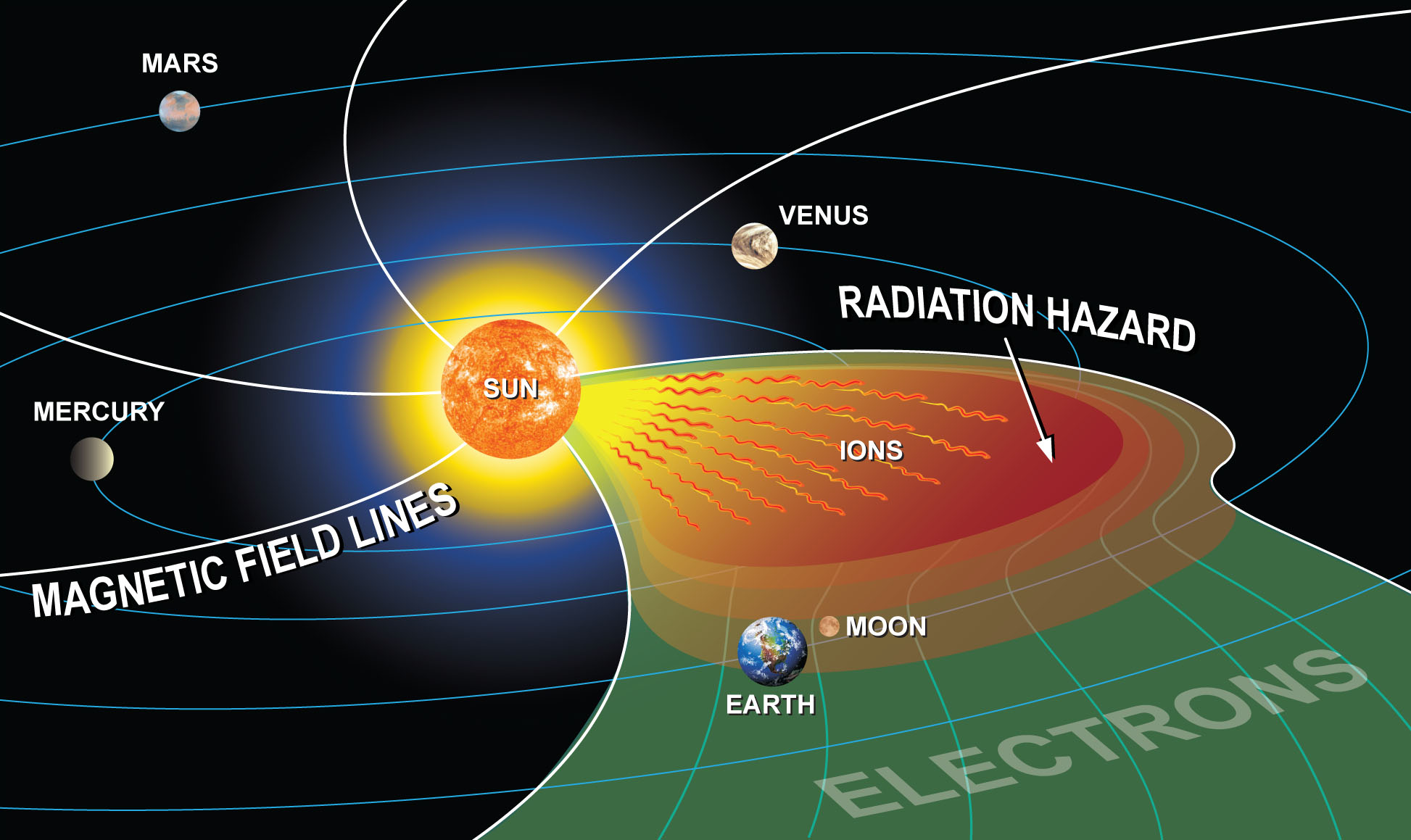}
     \caption{Example of Sun-Earth magnetical connection and accelerated
     particles movement following the Parker's spiral before reaching
     Earth. (Drawing courtesy from Space Weather) \cite{greendale}}
       \label{fig1}
\end{figure}

Finally, in the case of a CME, plasma and
magnetic field from the solar corona is released into the solar wind.
Though most of the solar particles have the same composition, they are labeled
differently depending on their energies starting from 1 keV, in the solar wind,
to more than 500 MeV. SEP events are composed of particles, predominantly
protons and electrons, with at least 1 MeV energy that last between 2-20 days
and have a range of fluxes of 5-6 orders of magnitude \cite{gabriel1996power}.
Only $>$100 MeV particles are discussed herein. It is generally accepted that
there are two types of SEP events, one is associated with CMEs and the other is
associated with flares that are called respectively gradual and impulsive
\cite{reames1999particle}.

In this paper, we propose a novel method for predicting SEP events $>$100 MeV
based on the proton and X-ray correlations time series using interpretable
decision tree models.
Predicting impulsive events is considered to be a more challenging problem than
predicting the gradual events that happen progressively and leave a large window
for issuing SEP warnings. While we are mainly concerned with impulsive events,
we used the gradual events as well to test our model with. The accelerated
impulsive events may or may not reach Earth depending on the location of their
parent event because their motion is confined by the magnetic field. More
specifically, in order for the accelerated particles to reach Earth, a Sun-Earth
magnetic connection needs to exist that allows the particles to flow to Earth
via the Parker spiral. Fig~.\ref{fig1} shows a cartoon of a solar eruption that
happened in the Western limb of the Sun that happens to be magnetically
connected to Earth.

Since SEP events are also part of solar activity, it may seem that
their occurrence is dependent on the solar cycle and therefore the number of
Sunspots on the Sun's surface, which is the case for other solar eruptions.
However, according to \cite{gabriel1996power}, there is no correlation between the solar
cycle and SEP event occurrence and fluences. In addition, there is no
evidence the dependence of SEP events on the number of Sunspots
that are present during that snapshot in time \cite{gabriel1996power}.
\par
 The rest
of the paper is presented as follows. In Section 2 we provide background
material on the SEP predictive models and related works. Section 3
defines our dataset used in this study, and then in Section 4 we lay out our
methodology. Finally, Section 5 contains our experimental results, and we finish
with conclusions and future work in Section 6.

\section{Related Works}
There are a number of predictive models of SEP events that can be categorized
into two classes: physics-based models \cite{p0, p1} and the precursor-based
models \cite{c0, c1}.
The first category of models includes the SOLar Particle Engineering Code
(SOLPENCO) that can predict the flux and fluence of the gradual SEP events
originating from CMEs \cite{aran2006solpenco}. However, such efforts mainly
focus on gradual events.
On the other hand, there are models that rely on historical observations to find
SEP events associated precursors. One example of such systems is the proton
prediction system (PPS), which is a program developed at the Air Force Research
Laboratory (AFRL), that predicts low energy SEP events E$>$\{5, 10, 50\} MeV,
composition, and intensities. PPS assumes that there is a relationship between
the proton flux and the parent flare event. PPS takes
advantage of the correlation between large SEP events observed by the
Interplanetary Monitoring Platform (IMP) satellites as well as their correlated
flare signatures captured by GOES proton, X-ray flux and H$\alpha$ flare
location \cite{PPS}. Also, the Relativistic Electron Alert System for
Exploration (RELEASE), predicts the intensity of SEP events using relativistic
near light speed electrons \cite{posner2007up}.
RELEASE uses electron flux data from the SOHO/COSTEP sensor of the range of
0.3-1.2 MeV to forecast the proton intensity of the energy range 30-50 MeV.
Another example of precursor-based models appear in
\cite{laurenza2009technique}, that base their study on the "big flare
syndrome". This latter theory states that SEP events occurrence at 1 AU is
highly probable when the intensity of the parent flare is high. Following this
assumption, the authors in \cite{laurenza2009technique} issue SEP forecasts for
important flares greater than M2. To this end, it uses type III radio burst
data, H$\alpha$ data, and GOES soft X-ray data. Finally, $GLE Alert Plus$, is an
operational system that uses a ground-based neutron
monitor (MNDB, www.nmdb.com) to issue alerts of SEP events of energies E$>$433
MeV. Finally, the University of Malaga Solar Energetic Particles Prediction
(UMASEP) is another system that first predicts whether a $>$10 MeV and $>$100
MeV SEP will happen or not. To do so, it computes the correlation between the
soft X-ray and proton channels to assess if there is a magnetic connection
between the Sun and Earth at the time of the X-ray event.
 Then, in case of existence of magnetic connection, UMASEP gives an
estimation on the time when the proton flux is expected to surpass the SWPC
threshold of J(E $>$ 10MeV)= 10$pfu$ and J(E $>$ 100MeV)= 1$pfu$ (1$pfu$ = $pr
cm^{-2} sr^{-1} s^{-1}$) and for the case of UMASEP-100, the intensity of the
first three hours after the event onset time.

 \begin{figure} 
    \centering
    \includegraphics[width=0.85\linewidth]{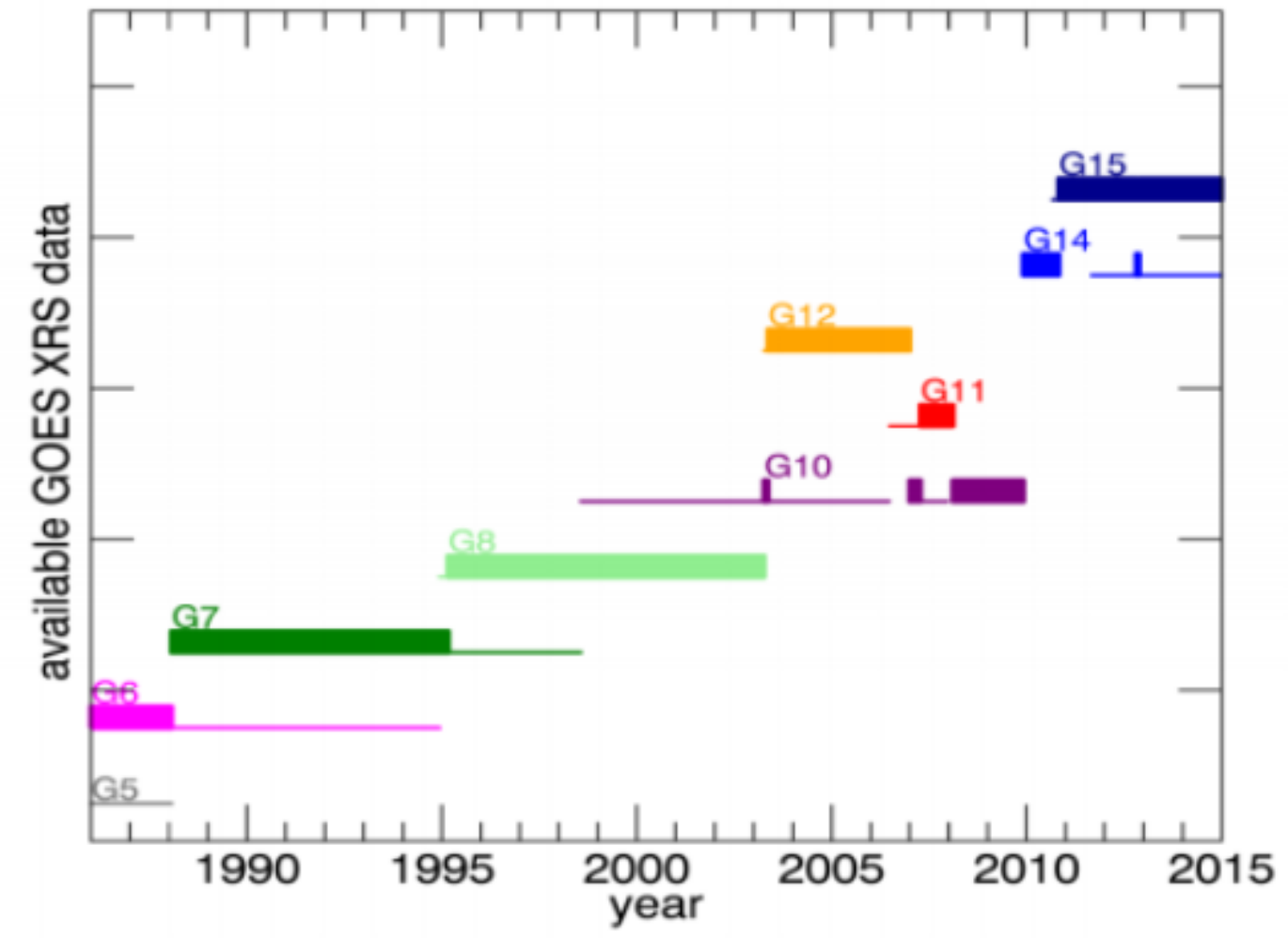}
     \caption{Primary (bold lines) and secondary (thin lines) GOES satellites
     for XRS data since 1986 (the primary and secondary satellites designation is
     unknown prior to 1986) (Figure from NOAA instrument report)}
       \label{fig2}
\end{figure} 

 \begin{figure} 
    \centering
    \includegraphics[width=1\linewidth]{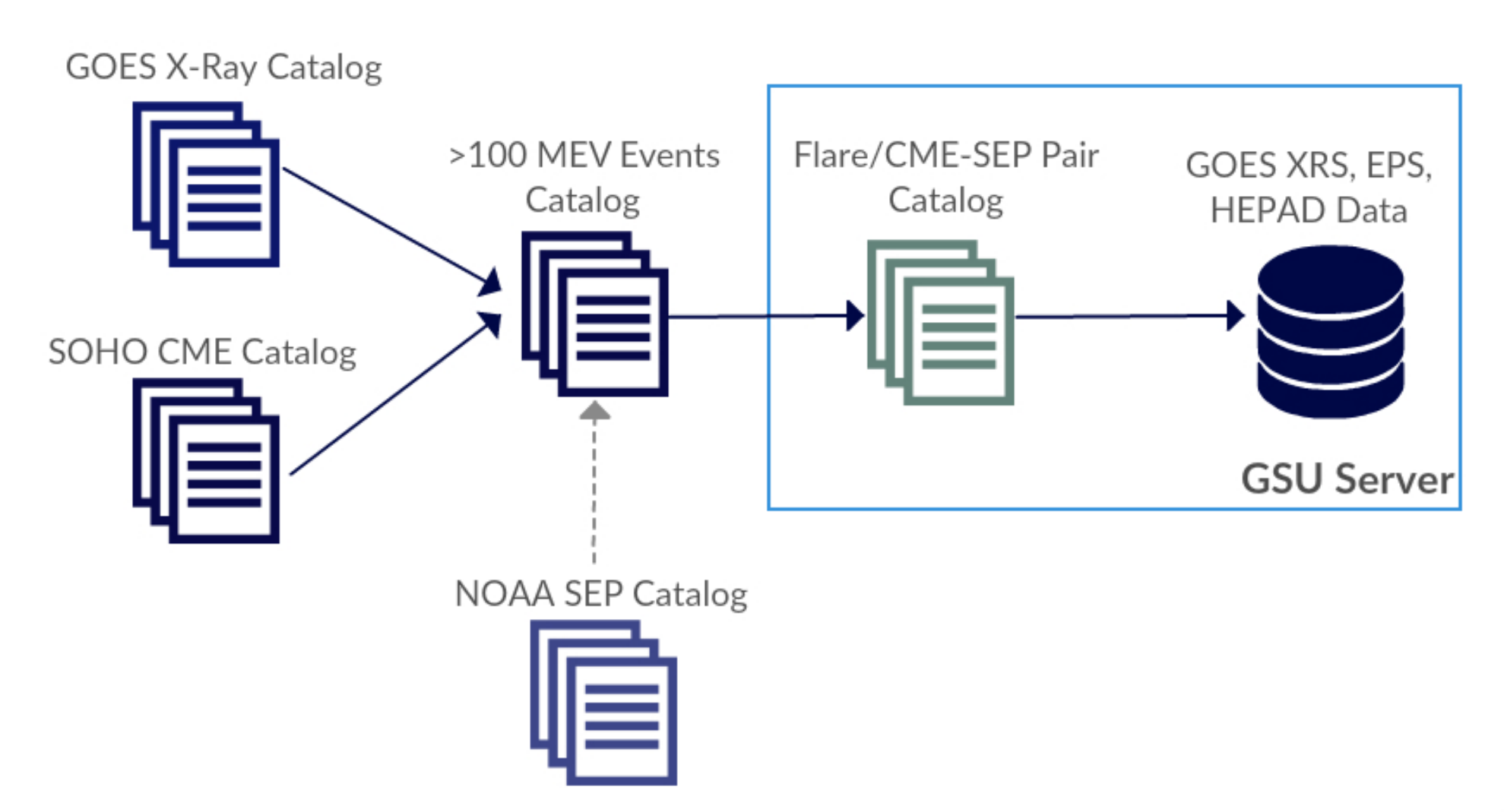}
     \caption{Catalogs used to make the x-ray-parent event mapping. X-ray and
     CME catalogs for detecting the parent event report for flare and CME
     respectively.}
       \label{cats}
\end{figure} 

While most of the SEP predictive systems either focus on the CME associated
events or low energy SEP events, with the exception of $GLE Alert Plus$ and
UMASEP, in this present work, we focus on higher energy SEP events that can be
more disruptive than lower energy events. In this work, we study the
GOES cross-channel correlations that can give an early insight on whether there
exist a magnetic connection or not.

We aim to provide an interpretable decision tree models using a
balanced dataset of SEP and non-SEP events.
The highest SEP energy band of $>$500 MeV or higher
that are measurable from the ground is out of the scope of this study.
Similarly, the lower SEP energy band of $<$100 MeV is not considered in this
study.

\section{Data}
Our dataset is composed of multivariate time series of X-ray, integral proton flux and fluences spectra that were measured on board of
Space Environment Monitor (SEM) instruments package of the Geostationary
Operational Earth Satellites (GOES).
In particular, we consider both the short and long X-ray channel data recorded
by the X-ray Sensor (XRS). For the proton channels, we consider channels P6 and
P7 recorded by the Energetic Particle Sensor (EPS) and proton channels P8, P9,
P10, and P11 recorded by the High Energy Proton and Alpha Detector (HEPAD).
Table~\ref{goesinstruments} summarizes the instruments onboard the GOES satellites
and their corresponding data channels that we used.

\begin{figure*}
\centering 
\begin{tabular}{ |c|c|c|}
Fast-Rising & Slow-Rising & Lack of SEP\\
\hline 
\includegraphics[width=0.315\linewidth]{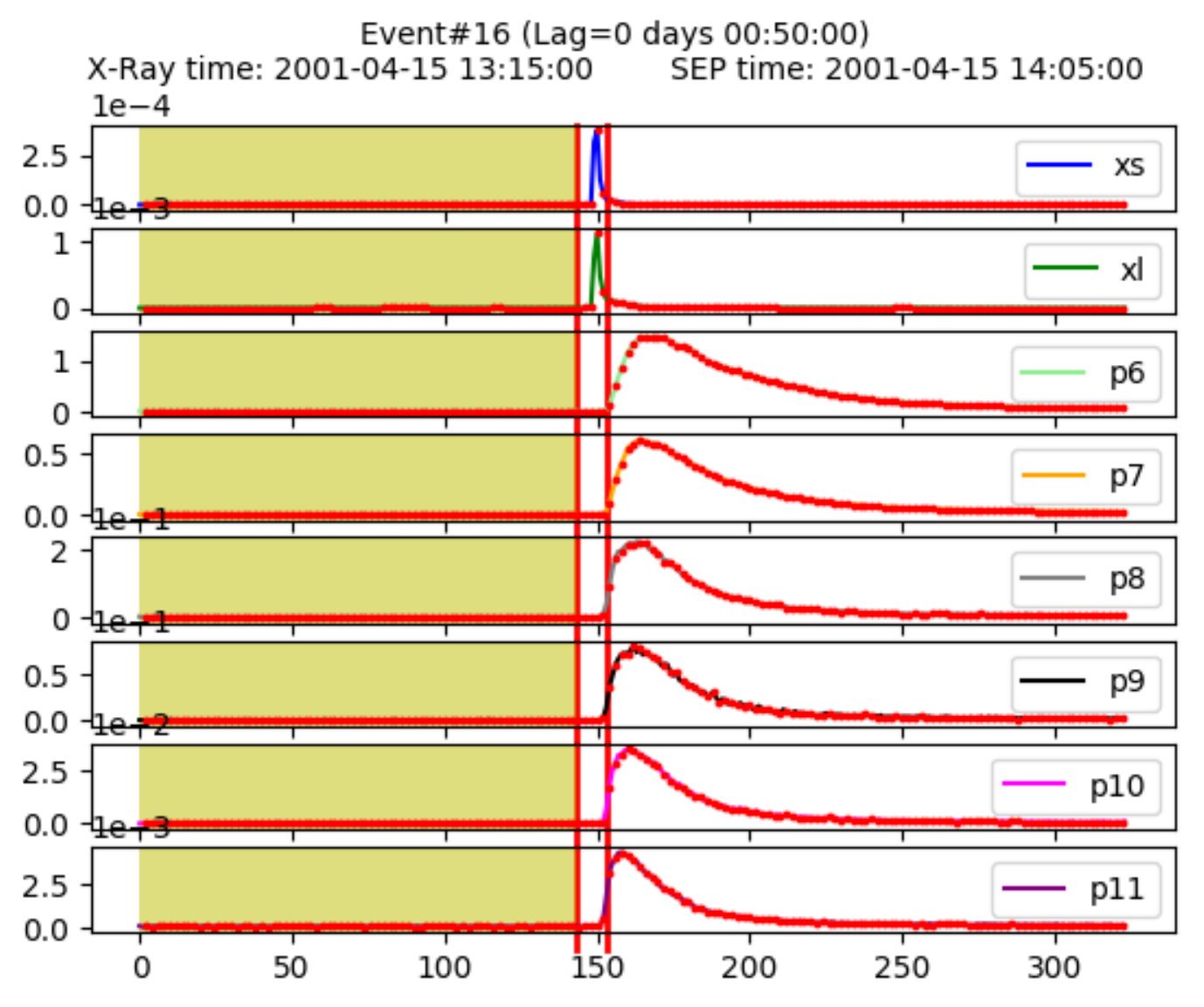}&
\includegraphics[width=0.32\linewidth]{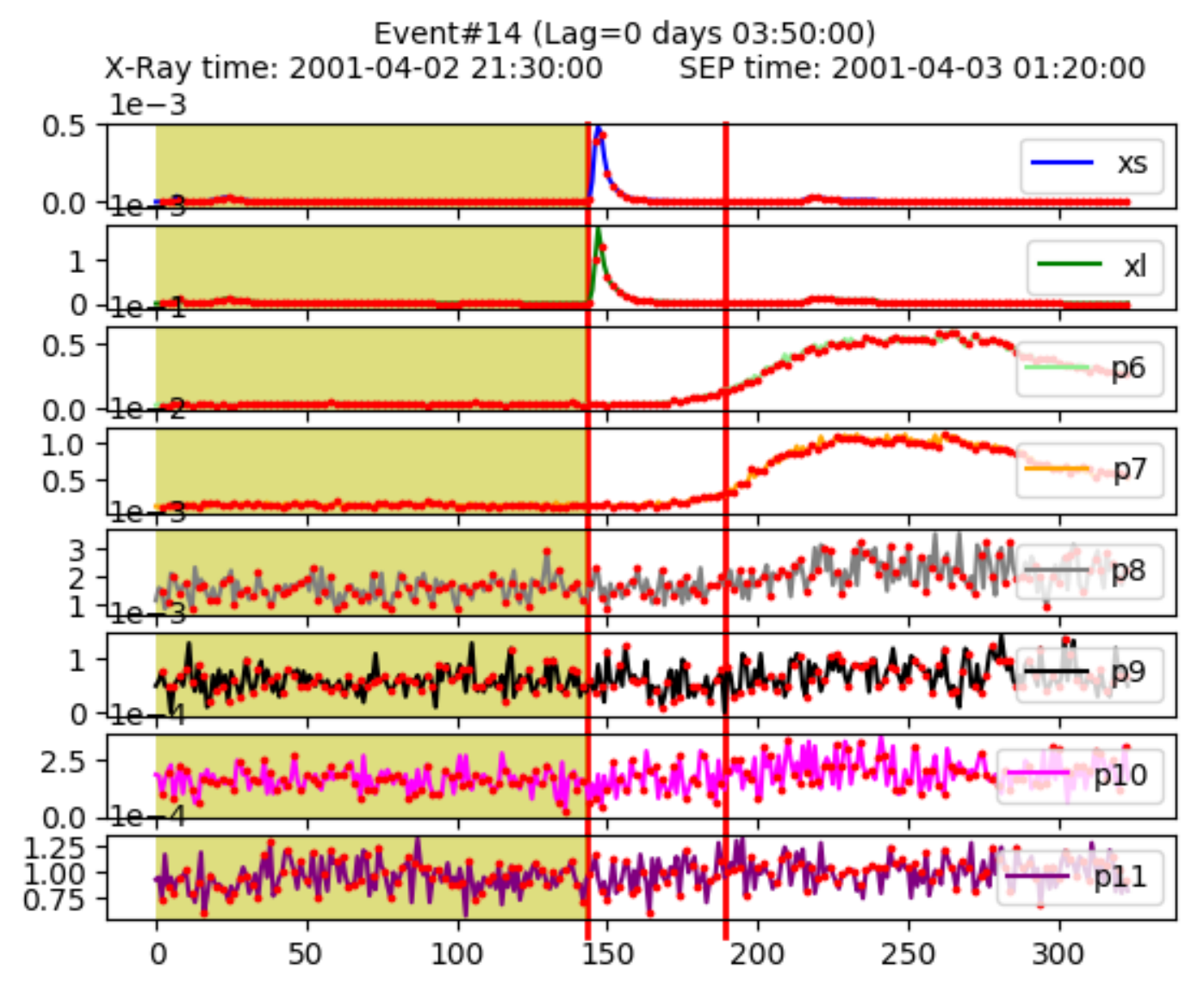}&
\includegraphics[width=0.32\linewidth]{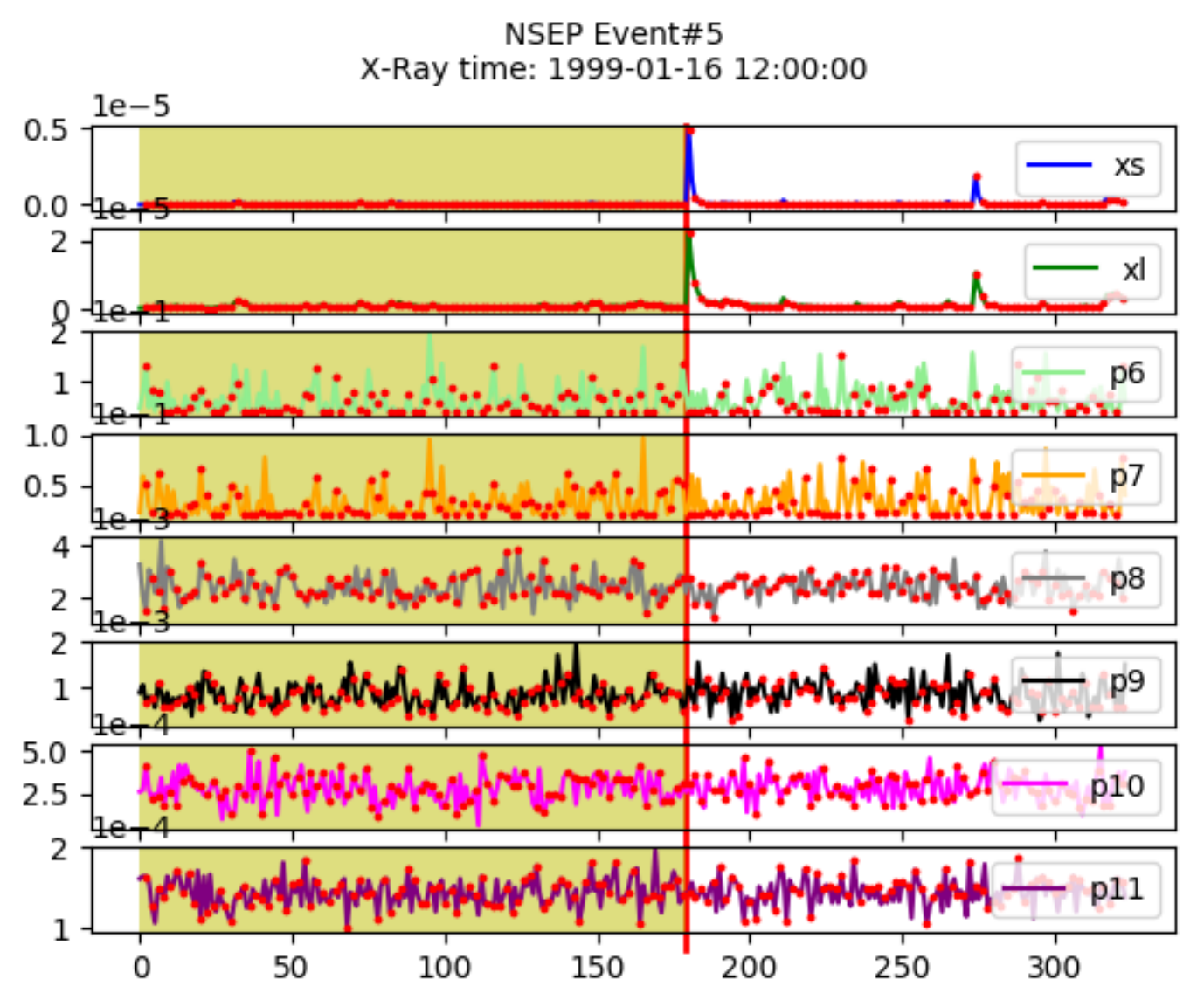}
\\
\hline
\includegraphics[width=0.175\linewidth]{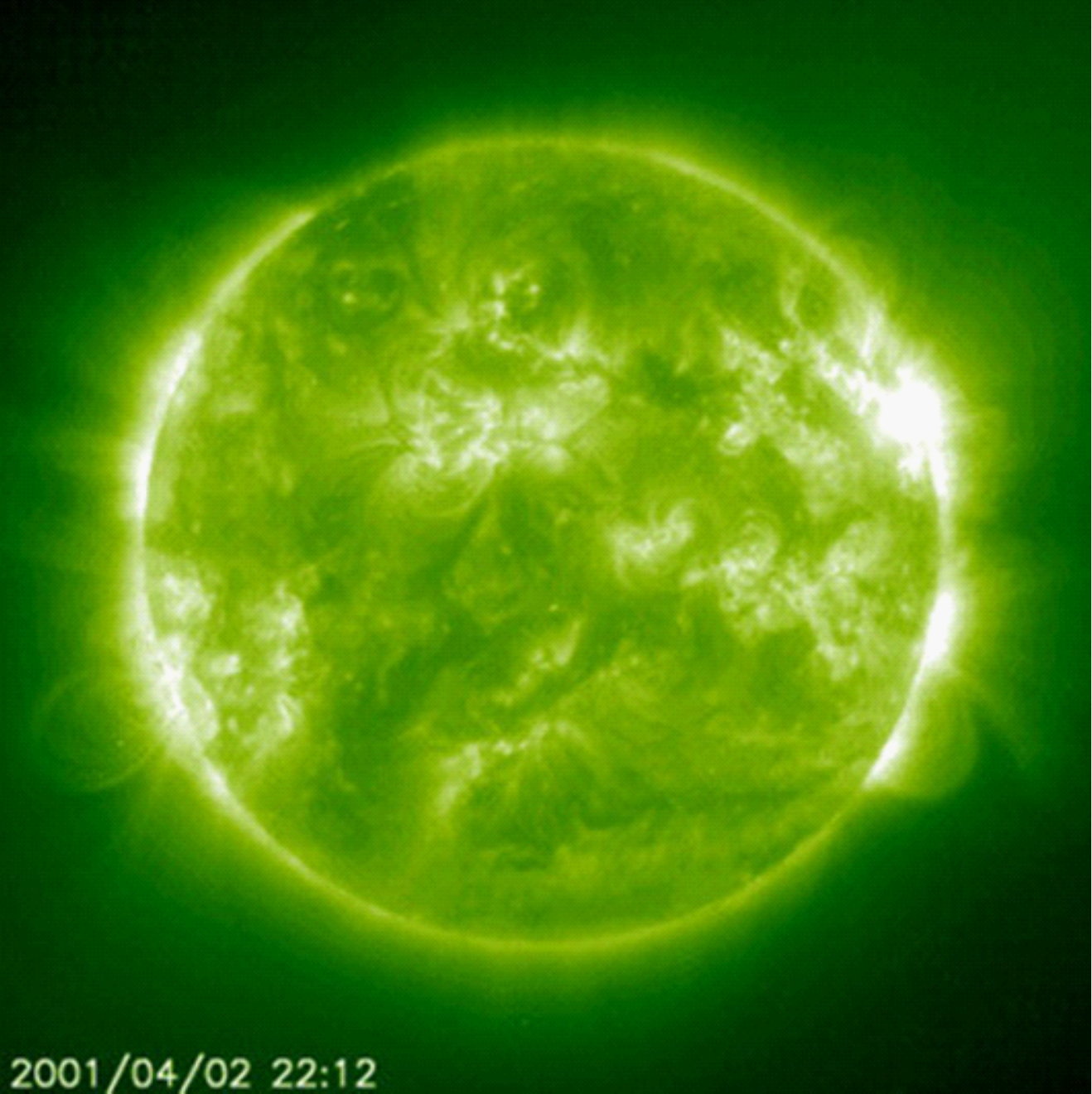} &
\includegraphics[width=0.17\linewidth]{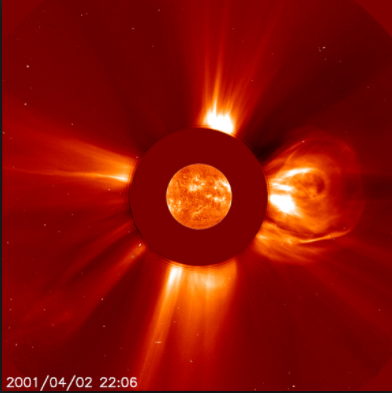}&
\includegraphics[width=0.17\linewidth]{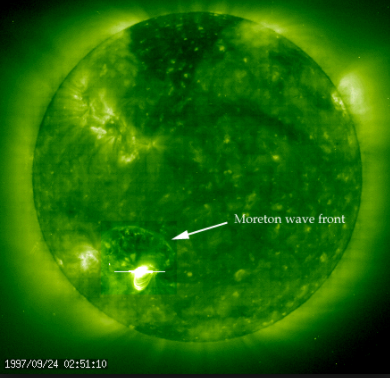}
\\
 \hline 
 \end{tabular}
 \caption{Example of an (a) Impulsive SEP event
     that started on the 2001-04-15 14:05:00 as a result of a flare
    that occurred in the 2011-04-15 13:15:00 shown in the SOHO EIT instument and
    a (b) gradual SEP event whose nearest temporal flare happened on 2001-04-02 21:30:03A and occurred as a
    result of a CME on the 2001-04-02 22:06:07 shown in the SOHO LASCO instrument
    and a (c) a flare that happened on 1999-01-16 12:00:00 that did not lead to any
    $>$ 100 MeV SEP event shown in the SOHO EIT
    instrument.\label{table1impulsive}}
\end{figure*}

 The data we collected is made publically available by NOAA in the
following link:
(\href{https://satdat.ngdc.noaa.gov/sem/goes/data/new\_avg/}{\textit{https://satdat.ngdc.noaa.gov/sem/goes/data/new\_avg/}}).
The data is available in three different cadences. The full resolution data is captured
every three seconds from the GOES satellites, which is aggregated and made
available with one and five minute cadences. In this paper we use the
aggregated five minute data which is the one usually cited in the literature
\cite{neal2001predicting} \cite{nunez2015real} \cite{nunez2011predicting}.
In most cases, there are a couple a co-existing GOES satellites whose data is
captured by more than one GOES satellite at a time. In this study, we always
consider the data reported by the primary GOES satellite that is designated by
the NOAA, as illustrated in Fig~.\ref{fig2}. The latter figure shows the primary GOES
satellite with a bold line and the other co-existing GOES for every year.
GOES-13 measurements were unstable for many years, but have been stable since
2014.

\begin{table*} 
\centering
\caption{GOES X-ray and Proton instuments and Channels.\label{goesinstruments}}
\begin{tabular}{ |c|c|c|c|}
\hline
Instrument & Channels & Description \\
\hline
\multirow{2}{*}{XRS}
 & xs &Short wavelength channel irradiance (0.5 - 0.3 nm)\\
 & xl &Long wavelength channel irradiance (0.1-0.8 nm)\\
  \hline
\multirow{4}{*}{HEPAD}
 & p8\_flux &Proton Channel 350.0 - 420.0 MeV \\
 & p9\_flux &Proton Channel 420.0 - 510.0 MeV \\
 & p10\_flux &Proton Channel 510.0 - 700.0 MeV \\
 & p11\_flux &Proton Channel $>$ 700.0 MeV \\
 \hline
 \multirow{2}{*}{EPS}
 & p6\_flux &Proton Channel 80.0 - 165.0 MeV \\
 & p7\_flux &Proton Channel 165.0 - 500.0 MeV \\
  \hline
  
 \end{tabular}
\end{table*}

\begin{table}
 \caption{$>$ 100 MeV SEP Event List with their Parent Events
(CME/Flare)\label{sepevents}}
\centering 
\begin{threeparttable}
\begin{tabular}{ |c|c|c|}
\hline 
SEP Event ID & Onset Time of SEP Event & Parent X-ray Event \\ \hline
1  &  1997-11-04 05:52:00 & 1997-11-04 05:52:00  \\ \hline
2  &  1997-11-06 11:49:00 & 1997-11-06 11:49:00  \\ \hline
3\tnote{*}  &  1998-04-20 09:38:00 & 1998-04-20 09:38:00  \\ \hline
4  &  1998-05-02 13:31:00 & 1998-05-02 13:31:00  \\ \hline
5  &  1998-05-06 07:58:00 & 1998-05-06 07:58:00  \\ \hline
6  &  1998-08-24 21:50:00 & 1998-08-24 21:50:00  \\ \hline
7  &  1998-09-30 13:50:00 & 1998-09-30 13:50:00  \\ \hline
8  &  1998-11-14 05:15:00 & 1998-11-14 06:05:00  \\ \hline
9  &  2000-06-10 16:40:00 & 2000-06-10 16:40:00  \\ \hline
10  &  2000-07-14 10:03:00 & 2000-07-14 10:03:00  \\ \hline
11  &  2000-11-08 22:42:00 & 2000-11-08 22:42:00  \\ \hline
12  &  2000-11-24 14:51:00 & 2000-11-24 14:51:00  \\ \hline
13\tnote{*}  &  2000-11-26 16:34:00 & 2000-11-26 16:34:00  \\ \hline
14\tnote{*}  &  2001-04-02 21:32:00 & 2001-04-02 21:32:00  \\ \hline
15  &  2001-04-12 09:39:00 & 2001-04-12 09:39:00  \\ \hline
16  &  2001-04-15 13:19:00 & 2001-04-15 13:19:00  \\ \hline
17  &  2001-04-17 21:18:00 & 2001-04-18 02:05:00  \\ \hline
18  &  2001-08-15 12:38:00 & 2001-08-16 23:30:00  \\ \hline
19\tnote{*}  &  2001-09-24 09:32:00 & 2001-09-24 09:32:00  \\ \hline
20  &  2001-11-04 16:03:00 & 2001-11-04 16:03:00  \\ \hline
21  &  2001-11-22 22:32:00 & 2001-11-22 19:45:00  \\ \hline
22  &  2001-12-26 04:32:00 & 2001-12-26 04:32:00  \\ \hline
23  &  2002-04-21 00:43:00 & 2002-04-21 00:43:00  \\ \hline
24  &  2002-08-22 01:47:00 & 2002-08-22 01:47:00  \\ \hline
25  &  2002-08-24 00:49:00 & 2002-08-24 00:49:00  \\ \hline
26  &  2003-10-28 09:51:00 & 2003-10-28 09:51:00  \\ \hline
27  &  2003-11-02 17:03:00 & 2003-11-02 17:03:00  \\ \hline
28\tnote{*}  &  2003-11-05 02:37:00 & 2003-11-05 02:37:00  \\ \hline
29\tnote{+}  &  2004-11-01 03:04:00 & 2004-11-01 03:04:00  \\ \hline
30\tnote{+}  &  2004-11-10 01:59:00 & 2004-11-10 01:59:00  \\ \hline
31\tnote{+}  &  2005-01-16 21:55:00 & 2005-01-17 08:00:00  \\ \hline
32\tnote{+}  &  2005-01-20 06:36:00 & 2005-01-20 06:36:00  \\ \hline
33\tnote{+}  &  2005-06-16 20:01:00 & 2005-06-16 20:01:00  \\ \hline
34\tnote{+}    \tnote{*}  &  2005-09-07 17:17:00 & 2005-09-07 17:17:00  \\ \hline
35\tnote{+}    \tnote{*}  &  2006-12-06 18:29:00 & 2006-12-06 18:29:00  \\ \hline
36\tnote{+}  &  2006-12-13 02:14:00 & 2006-12-13 02:14:00  \\ \hline
37\tnote{+}  &  2006-12-14 21:07:00 & 2006-12-14 21:07:00  \\ \hline
38  &  2011-06-07 06:16:00 & 2011-06-07 06:16:00  \\ \hline
39  &  2011-08-04 03:41:00 & 2011-08-04 03:41:00  \\ \hline
40  &  2011-08-09 07:48:00 & 2011-08-09 07:48:00  \\ \hline
41  &  2012-01-23 03:38:00 & 2012-01-23 03:38:00  \\ \hline
42\tnote{*}  &  2012-01-27 17:37:00 & 2012-01-27 17:37:00  \\ \hline
43  &  2012-03-07 01:05:00 & 2012-03-07 01:05:00  \\ \hline
44  &  2012-03-13 17:12:00 & 2012-03-13 17:12:00  \\ \hline
45  &  2012-05-17 01:25:00 & 2012-05-17 01:25:00  \\ \hline
46\tnote{*}  &  2013-04-11 06:55:00 & 2013-04-11 06:55:00  \\ \hline
47  &  2013-05-22 13:08:00 & 2013-05-22 13:08:00  \\ \hline
 \end{tabular}
   \begin{tablenotes}
  \item[*] Gradual Events.
  \item[+] Missing Data in P6 and P7.
  \end{tablenotes}
  \end{threeparttable}  
\end{table}

\begin{table}
 \caption{Non SEP Event List \label{ns}}
\centering 
\begin{threeparttable}
\begin{tabular}{|c|c|c|}
\hline
Non SEP Event ID & X-ray Event & Class \\ \hline
1 &  1997-09-24 02:43:00 &  M59  \\ \hline
2 &  1997-11-27 12:59:00 &  X26  \\ \hline
3 &  1997-11-28 04:53:00 &  M68  \\ \hline
4 &  1997-11-29 22:28:00 &  M64  \\ \hline
5 &  1998-07-14 12:51:00 &  M46  \\ \hline
6 &  1998-08-18 08:14:00 &  X28  \\ \hline
7 &  1998-08-18 22:10:00 &  X49  \\ \hline
8 &  1998-08-19 21:35:00 &  X39  \\ \hline
9 &  1998-11-28 04:54:00 &  X33  \\ \hline
10 &  1999-01-16 12:02:00 &  M36  \\ \hline
11 &  1999-04-03 22:56:00 &  M43  \\ \hline
12 &  1999-04-04 05:15:00 &  M54  \\ \hline
13 &  1999-05-03 05:36:00 &  M44  \\ \hline
14 &  1999-07-19 08:16:00 &  M58  \\ \hline
15 &  1999-07-29 19:31:00 &  M51  \\ \hline
16 &  1999-08-20 23:03:00 &  M98  \\ \hline
17 &  1999-08-21 16:30:00 &  M37  \\ \hline
18 &  1999-08-21 22:10:00 &  M59  \\ \hline
19 &  1999-08-25 01:32:00 &  M36  \\ \hline
20 &  1999-10-14 08:54:00 &  X18  \\ \hline
21 &  1999-11-14 07:54:00 &  M80  \\ \hline
22 &  1999-11-16 02:36:00 &  M38  \\ \hline
23 &  1999-11-17 09:47:00 &  M74  \\ \hline
24 &  1999-12-22 18:52:00 &  M53  \\ \hline
25 &  2000-01-18 17:07:00 &  M39  \\ \hline
26 &  2000-02-05 19:17:00 &  X12  \\ \hline
27 &  2000-03-12 23:30:00 &  M36  \\ \hline
28 &  2000-03-31 10:13:00 &  M41  \\ \hline
29 &  2000-04-15 10:09:00 &  M43  \\ \hline
30 &  2000-06-02 06:52:00 &  M41  \\ \hline
31 &  2000-06-02 18:48:00 &  M76  \\ \hline
32 &  2000-10-29 01:28:00 &  M44  \\ \hline
33 &  2000-12-27 15:30:00 &  M43  \\ \hline
34 &  2001-01-20 21:06:00 &  M77  \\ \hline
35 &  2001-03-28 11:21:00 &  M43  \\ \hline
36 &  2001-06-13 11:22:00 &  M78  \\ \hline
37 &  2001-06-23 00:10:00 &  M56  \\ \hline
38 &  2001-06-23 04:02:00 &  X12  \\ \hline
39\tnote{+}  &  2004-12-30 22:02:00 &  M42  \\ \hline
40\tnote{+}  &  2004-01-07 03:43:00 &  M45  \\ \hline
41\tnote{+}  &  2004-09-12 00:04:00 &  M48  \\ \hline
42\tnote{+}  &  2004-01-17 17:35:00 &  M50  \\ \hline
43\tnote{+}  &  2005-07-27 04:33:00 &  M37  \\ \hline
44\tnote{+}  &  2005-11-14 14:16:00 &  M39  \\ \hline
45\tnote{+}  &  2005-08-02 18:22:00 &  M42  \\ \hline
46\tnote{+}  &  2005-07-28 21:39:00 &  M48  \\ \hline
47\tnote{+}  &  2006-04-27 15:22:00 &  M79  \\ \hline
 \end{tabular}
  \begin{tablenotes}
  \item[+] Missing Data in P6 and P7.
  \end{tablenotes}
  \end{threeparttable} 
\end{table}


Only a portion of the collected data is used to train and test our
classifier. The positive class in this study is composed of X-Ray and proton
channels time series that led to $>$100 MeV SEP impulsive or gradual events. On
the other hand, the negative class is composed of X-Ray and proton channels
time series that did not lead to any $>$100 MeV SEP events. In order to
select such events we used a number of catalogs. For the positive class events
we used the same catalog of SEP events $>$100 MeV in
\cite{nunez2011predicting} that covers the events that happened between 1997 and
2013.

\par 

Our positive class is composed of the 47 X-Ray parent events of their
corresponding $>$100 MeV SEP events that appear in \cite{nunez2011predicting}
and shown in Table~\ref{sepevents}.
We use the X-Ray catalog
(\href{https://www.ngdc.noaa.gov/stp/space-weather/solar-data/solar-features/solar-flares/x-rays/goes/xrs/}{\textit{https://www.ngdc.noaa.gov/stp/space-weather/solar-data/solar-features/solar-flares/x-rays/goes/xrs/}})
as well as the CME catalog
(\href{https://cdaw.gsfc.nasa.gov/CME\_list/}{\textit{https://cdaw.gsfc.nasa.gov/CME\_list/}})
from the SOlar Heliospheric Observatory (SOHO) to derive the parent
events of the $>$100 MeV SEP events. There was an exception of two SEP events
that happened in August and September 1998 that we believe are gradual events but
could not map to any CME report due to the missing data during the SOHO mission
interruption. This latter happened because of the major loss of altitude
experienced by the spacecraft due to the failure to adequately monitor the
spacecraft status, and an erroneous decision which disabled part of the on-board
autonomous failure detection \cite{nunez2011predicting}.
It is worth to note that we consulted the NOAA-prepared
SEP events catalog along with their parent flare/CME events
(\href{ftp://ftp.swpc.noaa.gov/pub/indices/SPE.txt}{\textit{ftp://ftp.swpc.noaa.gov/pub/indices/SPE.txt}}).
For the case of events that are missing the NOAA catalog, we made our own
flare/CME-SEP events mapping. Fig.~\ref{cats} shows the three external catalogs
that we used to produce our own catalog from which we generate our SEP
dataset. 
To obtain a balanced dataset, we selected another 47 X-ray events that did not
produce any SEP events that is, shown in Table~\ref{ns}. We noticed
that there are nine SEP events (refer Table~\ref{sepevents}
ID:29-37) that happened during the period when only GOES-12
was operational as can be seen in Fig.~\ref{fig2}. At that period, channels P6
and P7 failed and there was no secondary GOES.
To make sure not to create any biased classifier that relies on the missing data to
make the prediction, we made sure to choose nine events from the negative class
as well that did not produce any SEP event (see Table~\ref{ns}
ID:39-47).

In this paper we make a clear distinction
between the two different classes of SEP events: gradual and impulsive. We
assume that an SEP event is flare accelerated, and therefore impulsive, if the lag
between the flare occurrence and the SEP onset time is very small and the peak
flux intensity has reached a global peak few minutes to an hour after the onset
time. On the other hand, a gradual event shows a progressive increase in the
proton flux trend that does not reach a global peak; instead, the peak is
maintained steadily before dropping again progressively. Finally, a non-SEP
event happens when there is an X-ray event of minimum intensity $M3.5$ that is
not followed by any significant proton flux increase in one of the P6-P11
channels. 

\section{Methodology}

This section introduces a novel approach in predicting the occurrence of $>$ 100
MeV SEP events based on interpretable decision tree models. We considered the
X-ray and proton channels as multivariate time series that entail some
correlations which may be precursors to the occurrence of an event. While \cite
{nunez2011predicting} considers the correlation between the X-ray and proton
channels only, we extended the correlation study into all the channels,
including correlations that happen across different proton channels. We
approached the problem from a multivariate time series classification
perspective. The classification task being whether the observed time series
windows will lead to an SEP event or not. There are two ways of performing a
time series classification. The first approach, which first appeared in
\cite{xi2006fast}, is to use the raw time series data and find the
K-nearest-neighbor with a similarity measures such as Euclidean distance,
and dynamic time warping. This approach is effective when the time series of
the same label shows a distinguishable similar shape pattern. In this problem,
the time series that we are working with are direct instruments readings that
show a jitter effect, which is common in electromechanical device readings
\cite{scargle1982studies}.
An example of the jitter effect is shown in P10, and P11 in
Figure.~\ref{table1impulsive}-b and Figure.~\ref{table1impulsive}-c. Time series
jitter makes it hard for distance measures, including elastic measures, to
capture similar shape patterns.
Therefore, we explored the second time series classification approach that
relies on extracting features from the raw time series before feeding it to a
model. In the next subsections, we will talk about the time series data
extraction, the feature generation and data pre-processing.

\subsection{Data Extraction}
Our approach starts from the assumption that a $>$100 MeV impulsive event may
occur if the parent X-ray event peak is at least $M3.5$ as was suggested
in \cite{nunez2011predicting}. Therefore we carefully picked the negative class
an X-ray event whose peak intensity is at least $M3.5$ but did not lead to any
SEP event (refer column 3 in Table~\ref{ns}). We extracted different
observation windows of data that we call a span. A span is defined as the number
of hours that constitute the observation period prior to an X-ray event. A total
of 94 (47*2) X-ray events (shown in column 3 and column 2 of
Table~.\ref{sepevents} and Table~.\ref{ns} respectively) were extracted
with different span windows. The span concept is illustrated in the yellow
shaded area in Figure.~\ref{table1impulsive}. The span window, in this case is
10 hours and stops exactly at the start time of the X-ray event. As we
considered the five minutes as the cadence between reports, a 10-hour span
window represents a 120-length multivariate X-ray and proton time series.

\subsection{Feature Generation}

To express the X-ray and proton cross-channel correlations we used a Vector
Autoregression Model (VAR) which is a stochastic process model used to capture
the linear interdependencies among multiple time series. VAR is the extension of
the univariate autoregressive model to multivariate time series. The VAR model
is useful for describing the behavior of economic, financial time series and for
forecasting \cite{zivot2006vector}. The VAR model permits us to express each
time series window as a linear function of past lags (values in the past) of
itself and of past lags of the other time series. The lag $l$ signifies the
factor by which we multiply a value of a time series
to produce its previous value in time.
Theoretically, if there exists a magnetic connection between the Sun and Earth
through the Parker spiral, the X-ray fluctuation precedes its corresponding
proton fluctuation. Therefore, we do not express the X-ray channels in terms of
the other time series, but, we focus on expressing the proton channels with
respect to the past lags of
themselves and with past lags of the X-ray channels (xs and xl). The VAR model
of order one, denoted as VAR(1) in our setting can be expressed
by Equations.(\ref{eq1})-(\ref{eq6}).
 
There is a total of eight time series that represent the proton channels. Every
equation highlights the relationship between the dependent variable and the
other protons and X-ray variables, which are independent variables. The higher
the dependence of a proton channel on an independent variable, the higher is
the magnitude of the coefficient $||\phi_{dependent\_{independent}}||$.
We used the coefficients of the proton equations
as a feature vector representing a data sample. The feature vector representing
a data point using the VAR(n) model is expressed in Equation.\ref{vec}.
  
Since the lag parameter $l$ determines the number of coefficients involved in the
equation, the number of features in the feature vector varies. More
specifically, the total number of features are 8 (independent variables) * 6 
(dependent variables).
\begin{table*}
\begin{equation}\label{eq1}
P6_{t,1} = \phi_{P6\_{xs,1}}*P6_{t-1,1} + \phi_{P6\_{xl,1}}*P6_{t-1,1} +
\phi_{P6\_{P6,1}}*P6_{t-1,1} + \ldots +
\phi_{P6\_{P11,1}}*P6_{t-1,1} + \alpha_{P6_{t,1}}
 \end{equation}
 \begin{equation}
P7_{t,1} = \phi_{P7\_{xs,1}}*P7_{t-1,1} + \phi_{P7\_{xl,1}}*P7_{t-1,1} +
\phi_{P7\_{P6,1}}*P7_{t-1,1} + \ldots +
\phi_{P7\_{P11,1}}*P7_{t-1,1} + \alpha_{P7_{t,1}}
 \end{equation}
 \begin{equation}
P8_{t,1} = \phi_{P8\_{xs,1}}*P8_{t-1,1} + \phi_{P8\_{xl,1}}*P8_{t-1,1} +
\phi_{P8\_{P6,1}}*P8_{t-1,1} + \ldots +
\phi_{P8\_{P11,1}}*P8_{t-1,1} + \alpha_{P8_{t,1}}
\end{equation}
\centering \vdots
\begin{equation}\tag{6}\label{eq6}
P11_{t,1} = \phi_{P11\_{xs,1}}*P11_{t-1,1} + \phi_{P11\_{xl,1}}*P11_{t-1,1} +
\phi_{P11\_{P6,1}}*P11_{t-1,1} + \ldots +
\phi_{P11\_{P11,1}}*P11_{t-1,1} + \alpha_{P11_{t,1}}
\end{equation}
\end{table*}

\begin{align}\tag{7}
    x &= \begin{bmatrix}
           \phi_{P6\_{xs,1}} \\
           \phi_{P6\_{xl,1}} \\
           \phi_{P6\_{P6,1}} \\
           \phi_{P6\_{P7,1}} \\
           \vdots \\
           \phi_{P11\_{P8,n}}\\ 
           \phi_{P11\_{P9,n}}\\ 
           \phi_{P11\_{P10,n}}\\ 
           \phi_{P11\_{P11,n}}\\ 
         \end{bmatrix} \label{vec}
  \end{align} 

\subsection{Data Preprocessing}
Before feeding the data to a classifier we cleaned the data from empty values
that appear in the generated features. To do so, we used the 3-nearest neighbors
class-level imputation technique. The method finds the 3 nearest neighbors
that have the same label of the sample with the missing feature. Nearest
neighbors imputation weights the samples using the mean squared difference on
features based on the other non-missing features. Then it imputes the missing
value with the nearest neighbor sample. The reason why
the imputation is done on a class level basis is that features may behave
differently across the two classes (SEP and non-SEP), therefore; it is
important to impute the missing data with the same class values.
%

\section{Experimental Evaluation}

In this section we explain the decision tree model that we will be using as well
as the sampling methodology. We will also provide a rationale for the choice of
parameters ($l$ and $span$). Finally we will zoom in the best model
with the most promising performance levels.

%
%
%
%

\subsection{Decision Tree Model}

A decision tree is a hierarchical tree structure used to determine classes based
on a series of rules/questions about the attribute values of the data points
\cite{safavian1991survey}.
Every non-leaf node represents an attribute split (question) and all the leaf
nodes represent the classification result. In short, given a set of features with
their corresponding classes a decision tree produces a sequence of
questions that can be used to recognize the class of a data sample.
In this paper, the data attributes are the VAR($l$) coefficients
$[\phi_{p6\_{xs, 1}}, \phi_{p6\_{xl, 1}}, ...,  \phi_{p6\_{xs, l}}]$ and the
classes are binary: SEP and non-SEP.

The decision tree classification model first starts by finding the variable that
maximizes the separation between classes.
Different algorithms use different metrics, also called purity measures, for
measuring the feature that maximizes the split. Some splitting criteria include
Gini impurity, information gain, and variance reduction. The commonality between
these metrics is that they all measure the homogeneity of a given feature with
respect to the classes. The metric is applied to each candidate feature to
measure the quality of the split, and the best feature is used. In this paper we
used the CART decision tree algorithm, as appeared in
\cite{loh2011classification} and \cite{steinberg2009cart}, with Gini and
information gain as the splitting criteria.

\subsection{Parameter Choice}

 \begin{figure*} 
    \centering
    \includegraphics[width=0.8\linewidth]{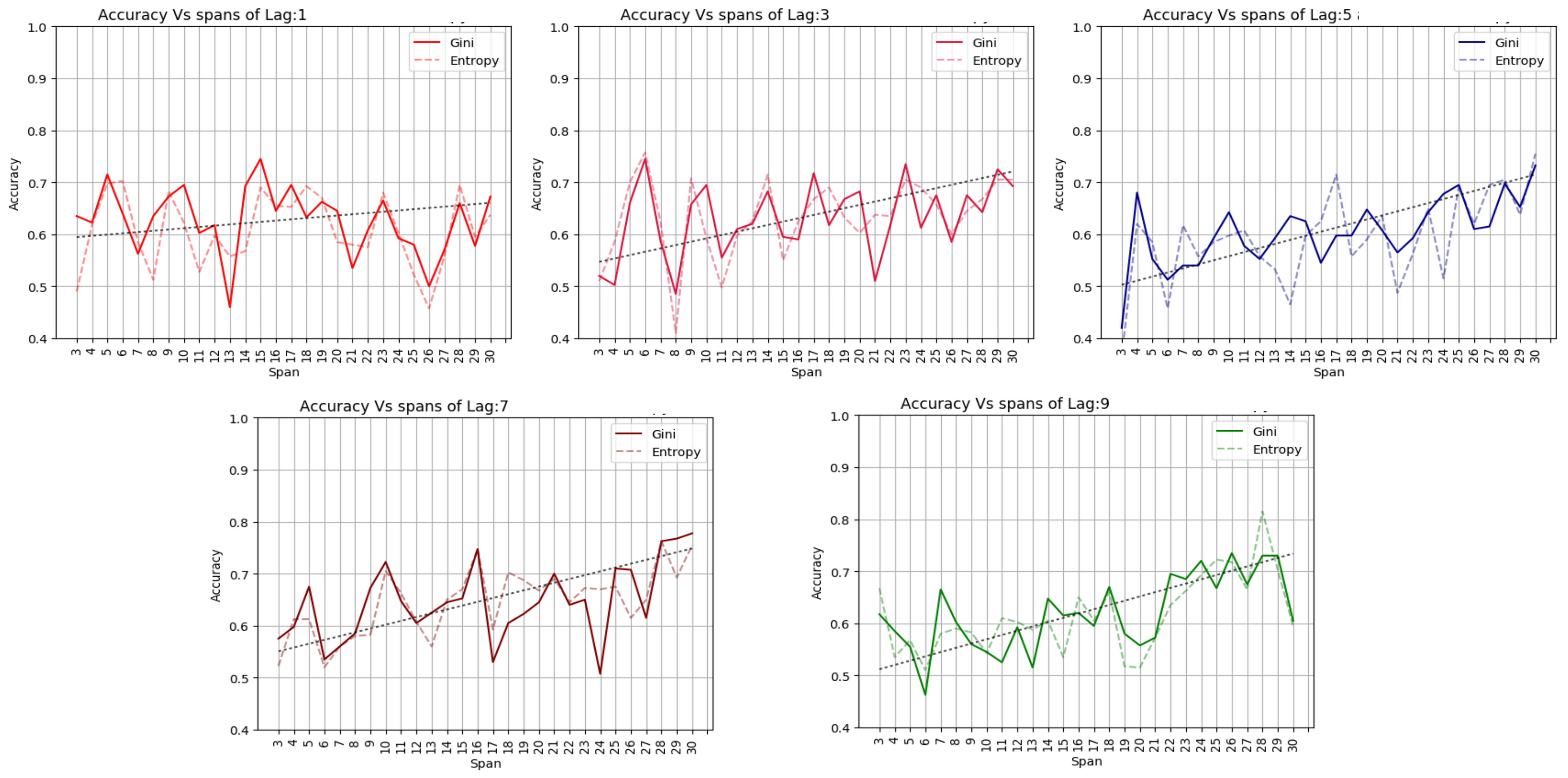}
     \caption{Decision tree accuracy with respect to the span window and the
     lag parameters using Gini and information gain splitting criteria. The
     dotted line shows a linear fit to the accuracy curve. }
       \label{spanacc}
\end{figure*} 

Our approach relies heavily on the choice of parameters, namely, the span window
and the VAR model lag parameter. The span is the number of observation hours
that precede the occurrence of an X-ray event. The latter determines the length
of the multivariate time series to be extracted. On the other hand, the lag
($l$) determines the size of the feature space that will be used as well as the
length of the dependence of the time series with each other in the past.
As mentioned previously, with a one-step increment of the lag parameter the size of the
feature space almost doubles $features\_number$ = 8*(independent variables)*6
(equations)*$l$+6*(equations). In order to determine the optimal parameters to
be used, we run a decision tree model on a set of values for both the span and
lag parameters. More specifically, we used the range [3-30] for the span window
and the set \{1,3,5,7,9\} for $l$. Since we have a balanced dataset we used a
stratified Ten-fold cross validation as the sampling methodology. A stratified
sampling always ensures a balance in the number of positive and negative samples for
both the training and testing data samples. Ten-fold cross-validation randomly
splits the data into 10 subsets, models are trained with nine of the
folds (90\% of the dataset), and test it with one fold (10\% of the dataset).
Every fold is used once for testing and nine times for training. In our
experiments, we report the average accuracy on the 10 folds.
 Fig.~\ref{spanacc} illustrates the accuracy curves with respect to the span
 windows for the five lags that we considered. We reported the accuracies of the
 decision tree model using both gini and information gain splitting criteria. In
 order to better capture the model behavior with the increasing span we
 plotted a linear fit to the accuracy curves of each lag. The first observation
 that can be made is that the slopes of the linear fit for $l$=1 and $l$=3 are
 relatively small in comparison to the other lags ($l>$3). This signifies that
 the model does not show any increasing or decreasing accuracy trend with the
 increase of the span window. Therefore we conclude that $l$=1 and $l$=3 are
 too small to discover any relationship between the proton and X-ray channels.
 Having the lag parameter set to $l$=1 and $l$=3 corresponds to expressing the
 time series (independent variable) going back in time up to five minute and 15
 minutes respectively. These latter times are small, especially for $l$=1 (5
 minutes), which theoretically is not possible since the
 protons can at most reach the speed of light that corresponds to a lag of at least 8.5
 minutes. For the other lags ($l>3$) there is noticeable increase in steepness in
 the accuracy linear fit which suggests that the accuracy increases with
 the increasing span window. The second observation is that for all the $l>3$
 datasets the best accuracy was achieved in the last four span window (i.e span
 $\in$ \{27,28,29,30\}). Therefore, we filtered the initial range of
 parameter values to \{5,7,9\} for $l$ and \{27,28,29,30\} for the span. In the
 next subsection we will zoom in into every classifier within the parameter
 grid.
 
 \subsection{Learning Curves}
\begin{figure*} 
 \centering 
 \begin{tabular}{ c  c  c  c  c }
  & Span 27 & Span 28 & Span 29 & Span 30 \\ 
 
 \begin{turn}{90} \hspace{10mm} Lag 5  \end{turn}&
 \includegraphics[width=0.221000001\linewidth]{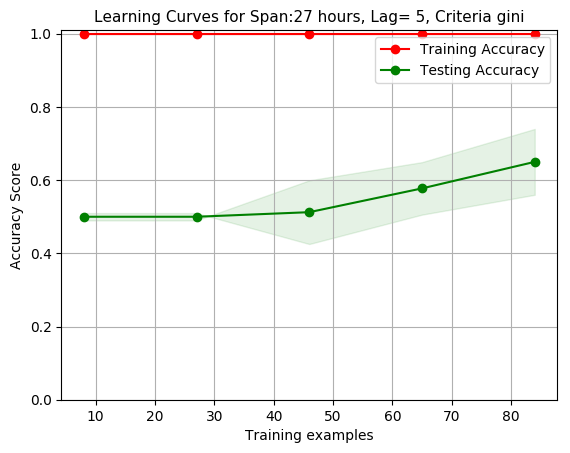} &
 \includegraphics[width=0.22\linewidth]{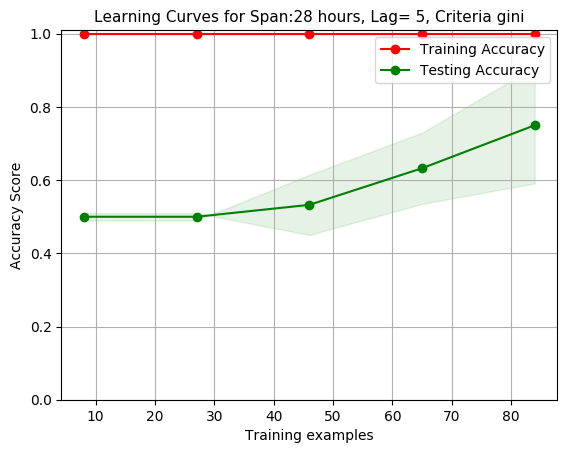} &
 \includegraphics[width=0.22\linewidth]{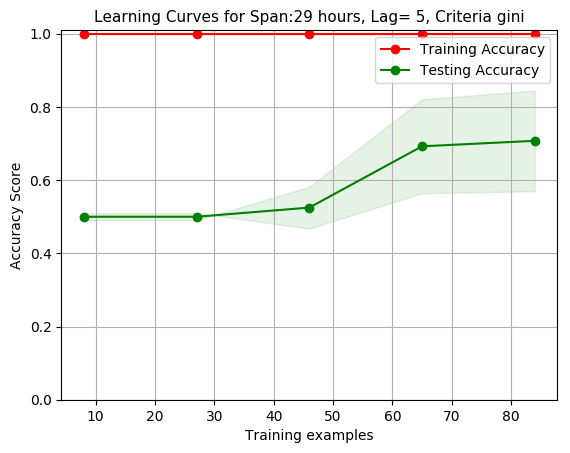}&
 \includegraphics[width=0.22\linewidth]{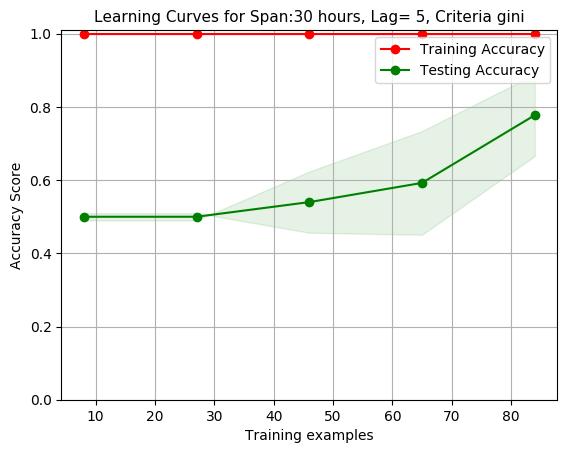}
  \\ 
 \begin{turn}{90} \hspace{10mm} Lag 7  \end{turn}&
 \includegraphics[width=0.221000001\linewidth]{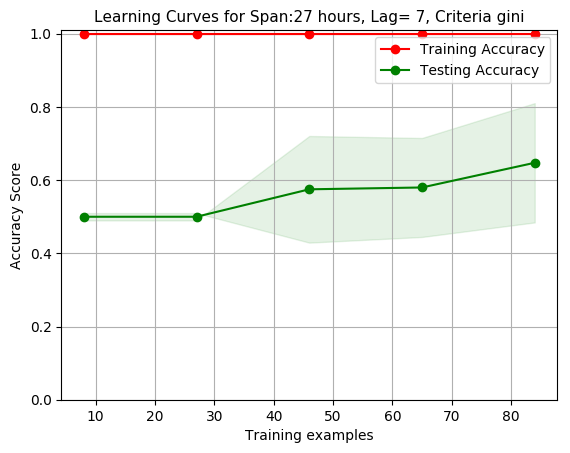} &
 \includegraphics[width=0.22\linewidth]{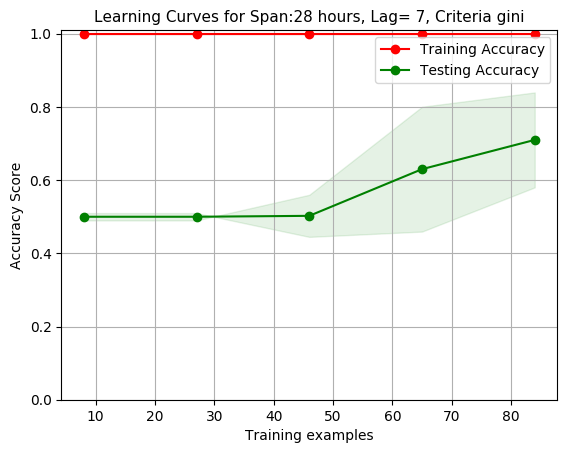} &
 \includegraphics[width=0.22\linewidth]{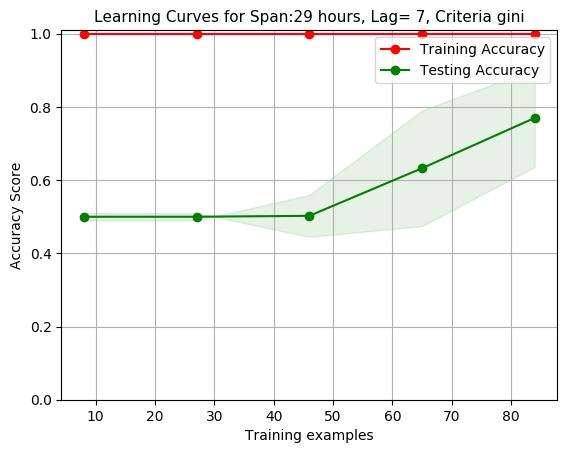}&
 \includegraphics[width=0.22\linewidth]{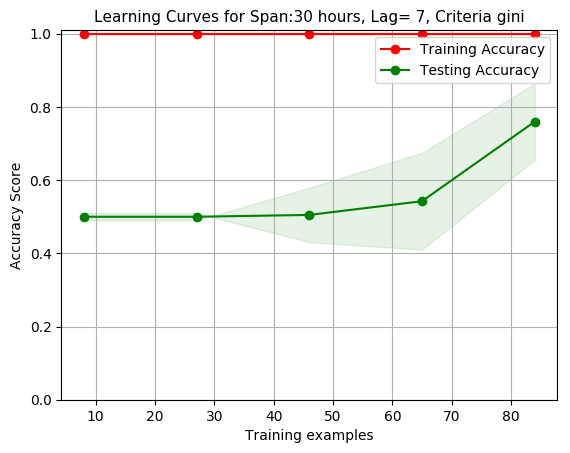}
  \\ 
 
 \begin{turn}{90} \hspace{10mm} Lag 9  \end{turn}&
 \includegraphics[width=0.221000001\linewidth]{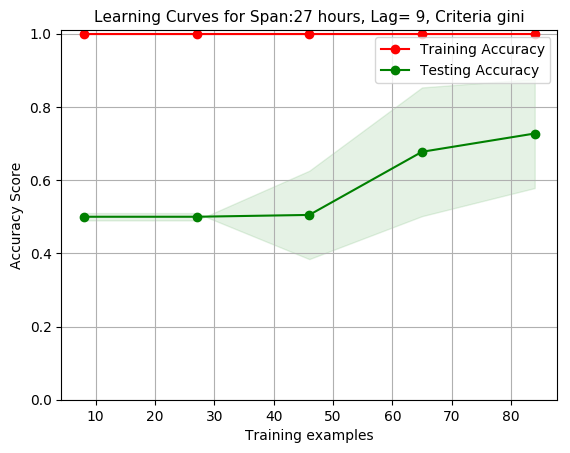} &
 \includegraphics[width=0.22\linewidth]{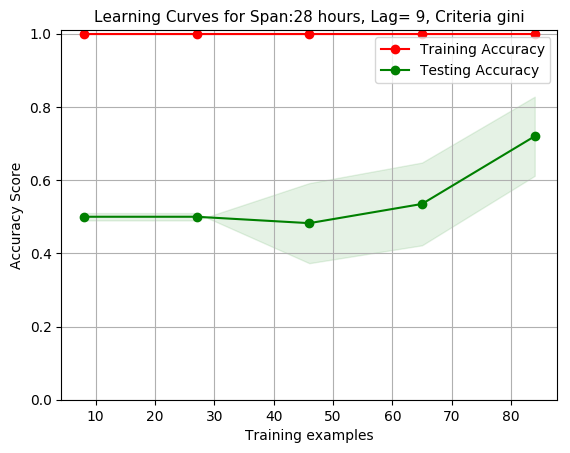} &
 \includegraphics[width=0.22\linewidth]{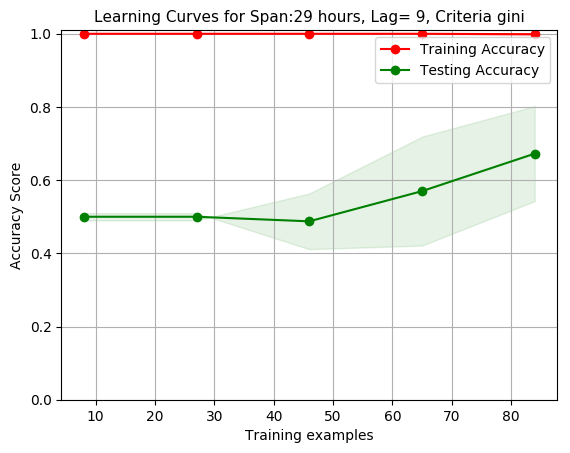}&
 \includegraphics[width=0.22\linewidth]{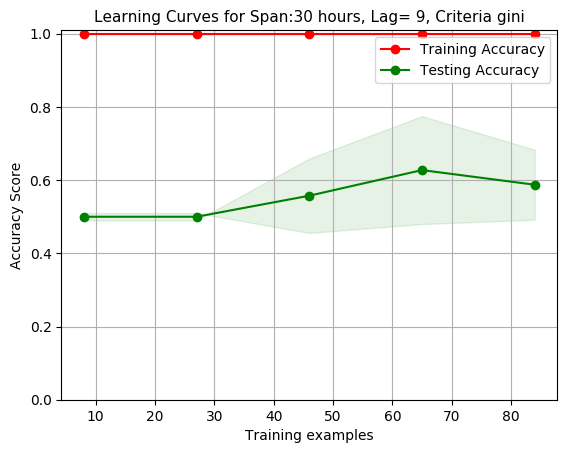}
  \\ 
  \end{tabular}
 	\caption{Learning curve of CART Decision Tree Models with Gini splitting
 criterion,spans $\in$ \{27,28,29,30\} and lag $\in$ \{5,7,9\}\label{lcgini}}
 
\end{figure*}

\begin{figure*} 
 \centering 
 \begin{tabular}{ c  c  c  c  c }
  & Span 27 & Span 28 & Span 29 & Span 30 \\ 
 
 \begin{turn}{90} \hspace{10mm}  Lag 5  \end{turn}&
 \includegraphics[width=0.221000001\linewidth]{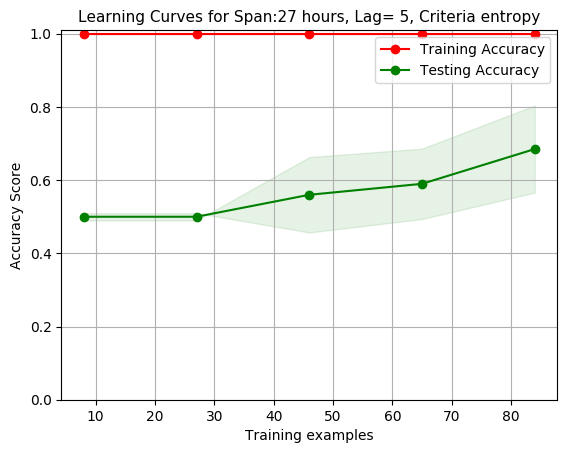} &
 \includegraphics[width=0.22\linewidth]{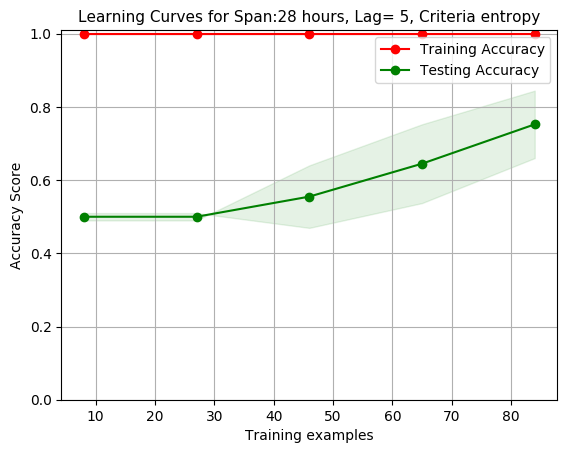} &
 \includegraphics[width=0.22\linewidth]{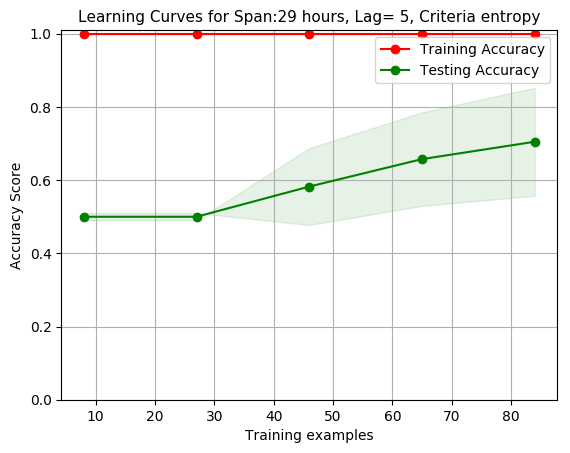}&
 \includegraphics[width=0.22\linewidth]{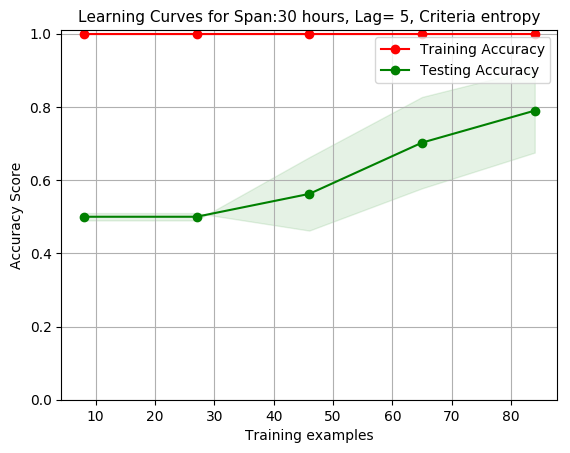}
  \\ 
 \begin{turn}{90} \hspace{10mm}  Lag 7  \end{turn}&
 \includegraphics[width=0.221000001\linewidth]{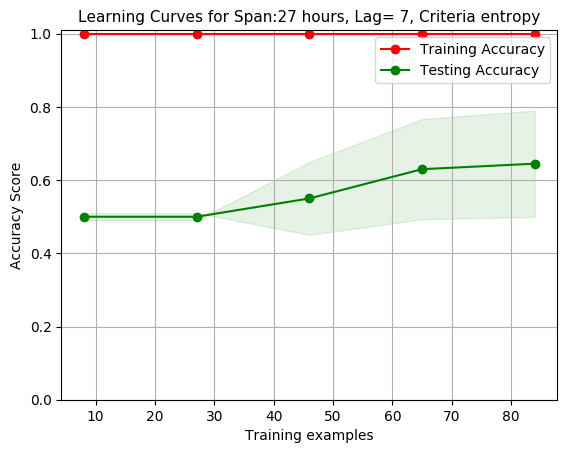} &
 \includegraphics[width=0.22\linewidth]{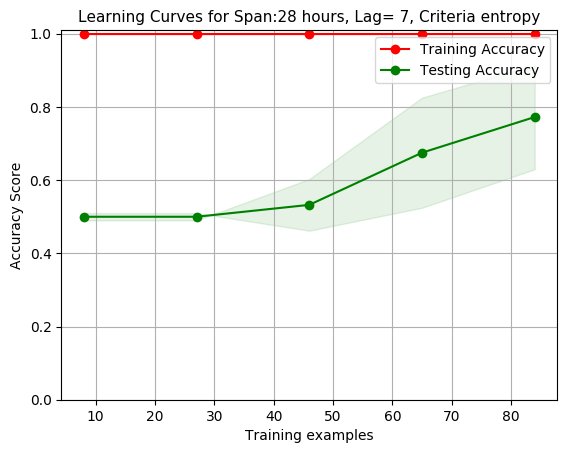} &
 \includegraphics[width=0.22\linewidth]{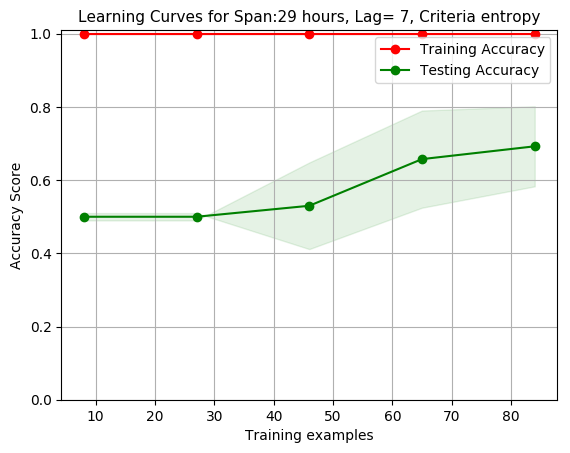}&
 \includegraphics[width=0.22\linewidth]{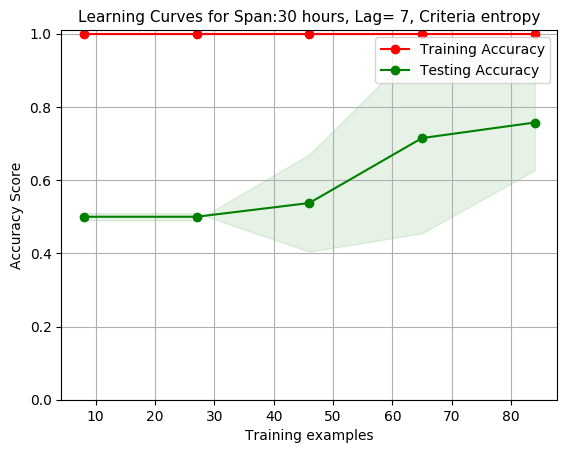}
  \\ 
 
 \begin{turn}{90} \hspace{10mm}  Lag 9  \end{turn}&
 \includegraphics[width=0.221000001\linewidth]{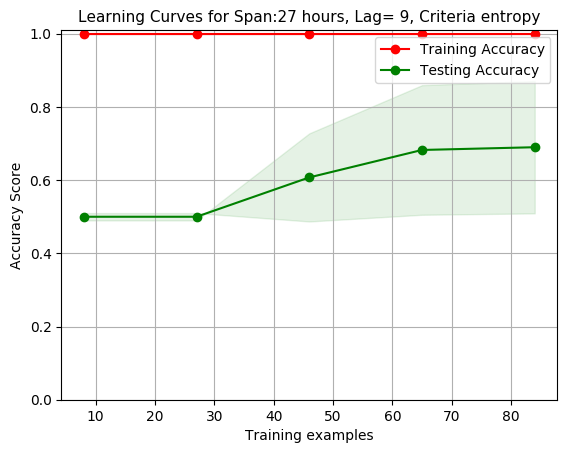} &
 \includegraphics[width=0.22\linewidth]{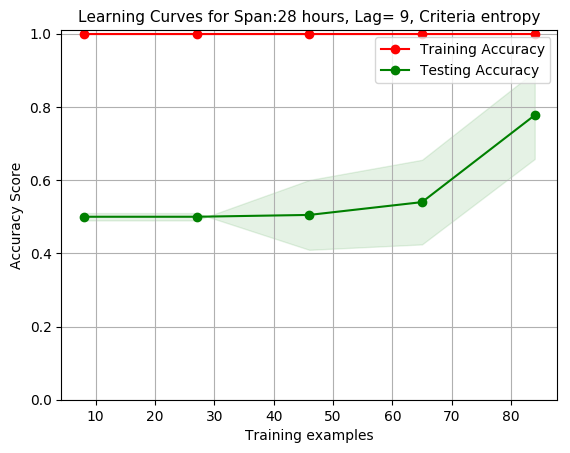} &
 \includegraphics[width=0.22\linewidth]{figs/9s29_gini_m.png}&
 \includegraphics[width=0.22\linewidth]{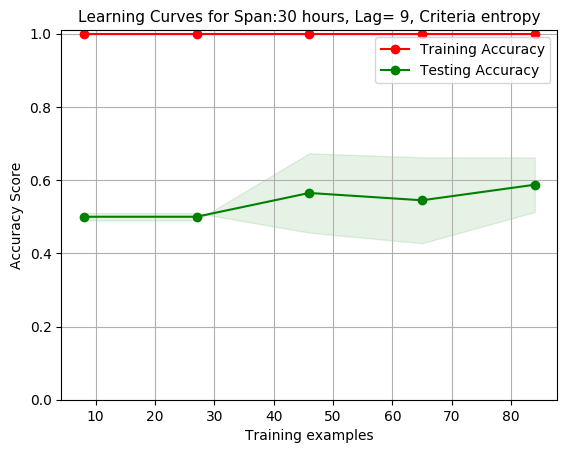}
  \\ 
 
  \end{tabular}
   \caption{Learning curve of CART Decision Tree Models with information gain
   splitting criterion,spans $\in$ \{27,28,29,30\} and lag $\in$ \{5,7,9\}\label{lcentropy} }
\end{figure*}

To be able to discriminate decision tree models that show similar
accuracies we use the model learning curves, also called experience curves, to
have an insight in how the accuracy changes as we feed the model with more
training examples.
Learning curves are particularly useful for comparing different
algorithms \cite{madhavan1997new} and choosing optimal model parameters during
the design \cite{pedregosa2011scikit}. It is also a good tool for visually
inspecting the sanity of the model in case of overtraining or undertraining.
Figs.~\ref{lcgini} and \ref{lcentropy} show the learning curves of the
decision tree model using gini and information gain as the splitting criteria
respectively. The red line represents the training accuracy which evaluates the
model on the newly trained data. The green line shows the testing
accuracy which evaluates the model on the the never-seen testing data. The
shaded area represents the standard deviation of the accuracies after running
the model multiple times with the same number of training data. It is noticeable
that the standard deviation becomes higher as the lag is increased. Also, it can
be seen that the best average accuracies, that appeared in Fig.~\ref{spanacc}, are
not always the ones that have the best learning curves. For example from
Fig.~\ref{spanacc}, the best accuracy that has been reached appears to be in
$l$=7 and $span=27,29$; however, the learning curves corresponding to that span
and lag show that the standard deviation is not very smooth as compared to
$l$=5. Therefore the experiments show that using $l$=5 results in relatively
stable models with lower variance. Therefore, we will
zoom in $l$=5 for all the spans $\in \{27,28,29,30\}$ that we previously
filtered.

\begin{table}[]
\centering
\caption{Decision Tree model evaluation for gini and information gain splitting
criteria\label{eval}}
\label{my-label}
\begin{tabular}{l|l|l|l|l|l|l|l|l|}
\cline{2-9}
                                & \multicolumn{4}{c|}{Gini}          & \multicolumn{4}{c|}{Information Gain} \\ \hline
\multicolumn{1}{|l|}{Span}      & 27   & 28   & 29   & 30            & 27    & 28    & 29    & 30            \\ \hline
\multicolumn{1}{|l|}{Accuracy}  & 0.64 & 0.74 & 0.73 & \textbf{0.74} & 0.77  &0.70   & 0.67  & \textbf{0.78} \\ \hline
\multicolumn{1}{|l|}{Recall}    & 0.69 & 0.70  & 0.74 & \textbf{0.70}  & 0.76 &0.70  & 0.70   & \textbf{0.73} \\ \hline
 \multicolumn{1}{|l|}{Precision} & 0.62 & 0.75 & 0.75 & \textbf{0.76} & 0.78  & 0.72  & 0.72  & \textbf{0.86} \\ \hline
  \multicolumn{1}{|l|}{F1}        & 0.65 & 0.75 & 0.75 & \textbf{0.74} & 0.79  &0.71  & 0.69  & \textbf{0.82} \\ \hline
 \multicolumn{1}{|l|}{AUC}       & 0.65& 0.72 & 0.74 & \textbf{0.73} & 0.76  & 0.70   & 0.69  & \textbf{0.77} \\ \hline

\end{tabular}
\end{table}

 \begin{figure*} 
    \centering
    \includegraphics[width=0.8\linewidth]{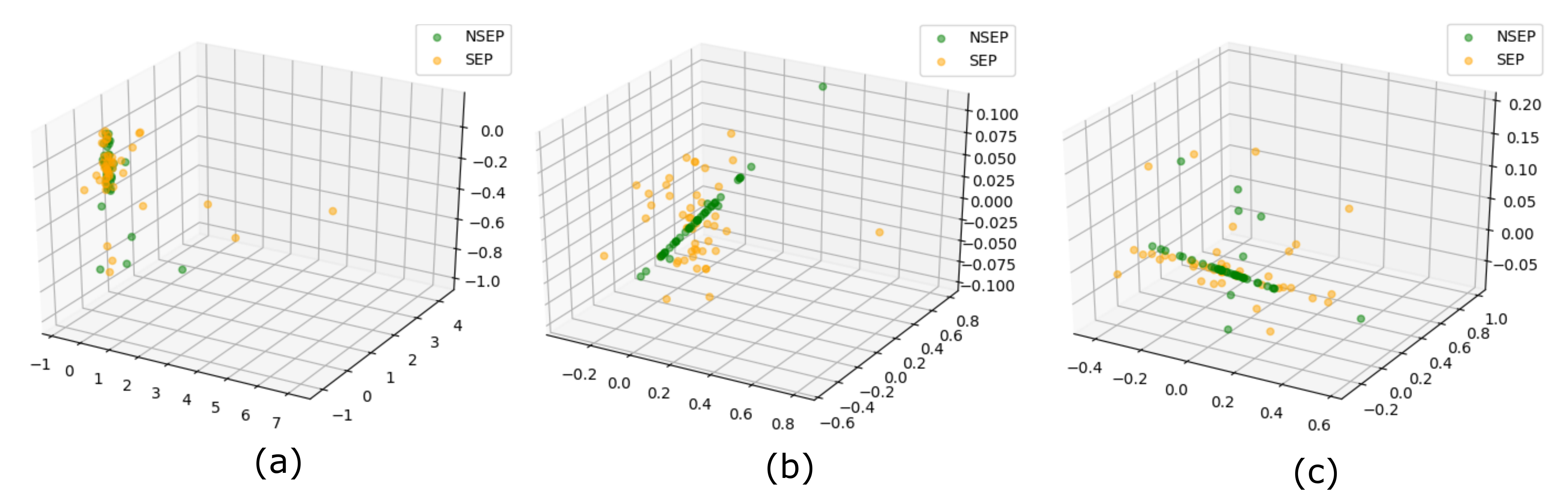}
     \caption{First 3 PCA components derived from (a) all the original 254 features, (b)
     the data sub-space containing only 4 parameters selected as the most
     relevant by the Gini index (as shown in the tree presented in Fig. 9),
     and (c) another data sub-space containing 4 different parameters (with 1
     repetition) selected as the most relevant by the Entropy measure. The
     PCA-based visualizations represent (sub-)spaces of the same data set (as
     shown in the tree presented in Fig. 10), with lag=5, and span=30.}
     
       \label{pca}
\end{figure*}

 \begin{figure} 
    \centering
    \includegraphics[width=1\linewidth]{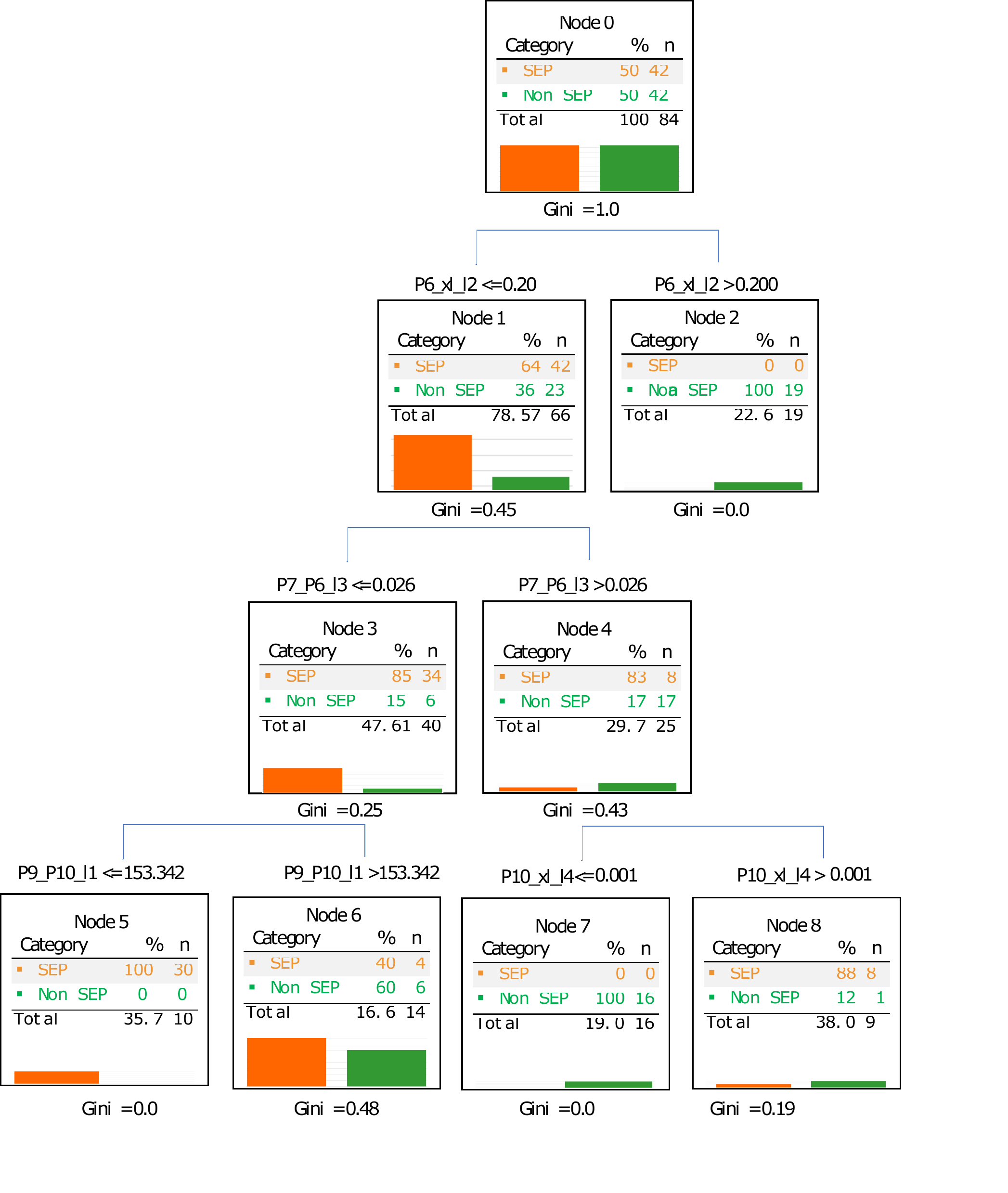}
     \caption{Decision Tree with Gini splitting criteria ($span$=30, $l$=5) }
       \label{giniDT}
\end{figure}  

 \begin{figure} 
    \centering
    \includegraphics[width=1\linewidth]{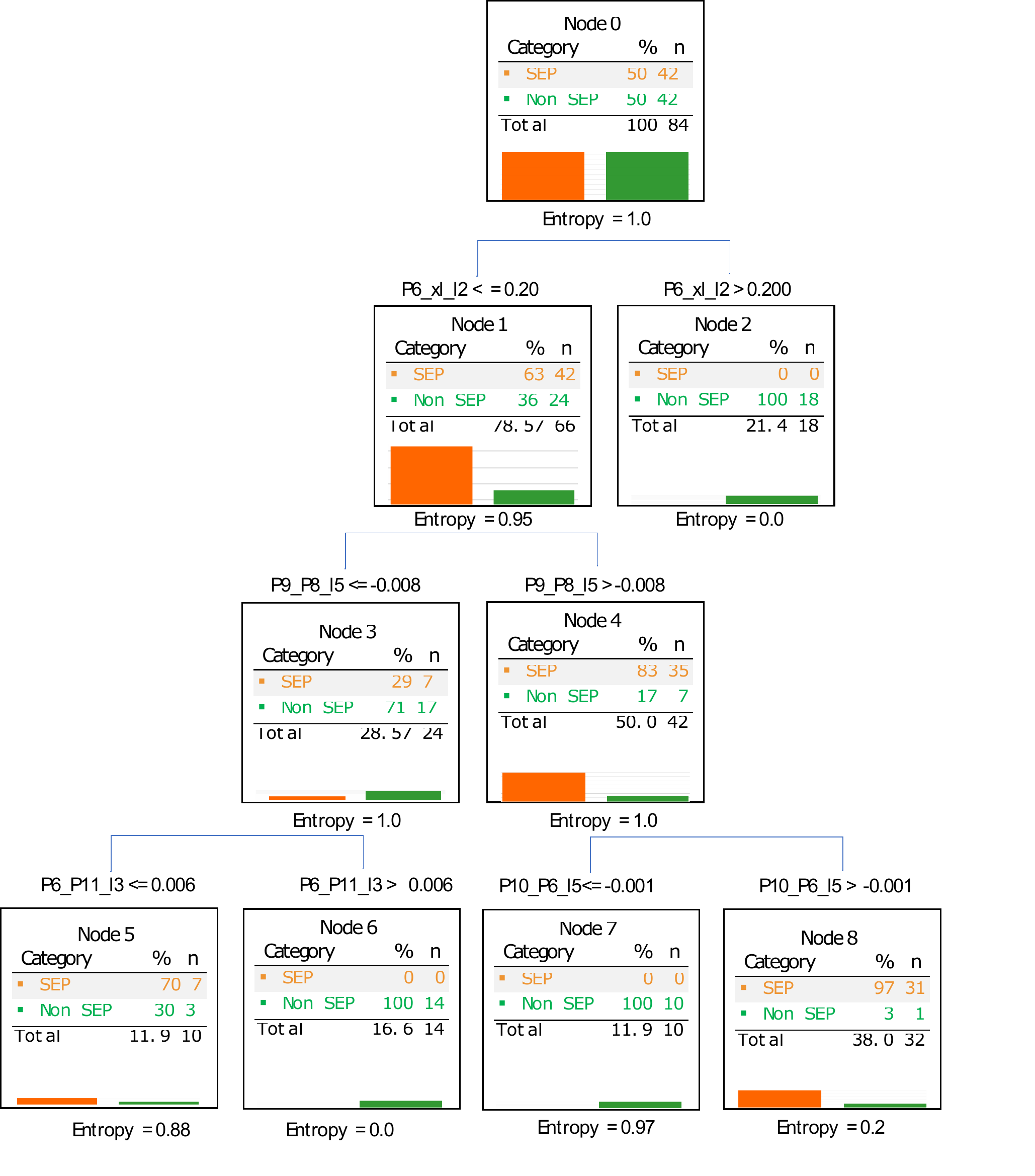}
     \caption{Decision Tree with information gain splitting criteria
     ($span$=30, $l$=5) }
       \label{entropyDT}
\end{figure}

To determine the best behaving model we choose six evaluation metrics that will
assess the models' performance from different aspects. Accuracy is the most
standard evaluation measure used to assess the quality of a classifier by
counting the ratio of correct classification over all the classifications.
In this context the accuracy measure is particularly useful because our
training and testing data is balanced. The data balance ensures that if the
classifier is highly biased toward a given class it will be reflected on the
accuracy measure. Recall is the second evaluation measure we considered, also
known as the probability of detection, which characterizes the
ability of the classifier to find all of the positive cases. Precision is
used to evaluate the model with respect to the false alarms. In fact, precision
is 1 - false alarm ratio. Precision and recall are usually anti-correlated;
therefore, a useful quantity to compute is their harmonic mean, the F1 score.
The last evaluation measure that we consider in the Area Under Curve (AUC) of the Receiver
Operating Characteristic curve (ROC) curve. The intuition behind this measure is
that AUC equals the probability that a randomly chosen positive example ranks
above (is deemed to have a higher probability of being positive than) a randomly
chosen negative example. It has been claimed in \cite{auccp} that the AUC is
statistically consistent and more discriminating than accuracy.

Table~.\ref{eval} shows the aforementioned evaluations on the $l$=5 datasets. It
is noticeable that span=30 achieves the best performance levels for both
splitting criteria. The decision tree models corresponding to those settings
using gini and information gain are shown in Fig.~\ref{giniDT} and
Fig.~\ref{entropyDT} respectively. For the purpose of visualization we used PCA
dimensionality reduction technique to plot the full feature space with the 254 dimensions of
the lag 5 and span 30 in Fig.~\ref{pca}-a, as well as the reduced feature space
with only the selected features from the gini measure in Fig.~\ref{pca}-b and
entropy measure in Fig.~\ref{pca}-c \cite{PCA}.
It is clearly visible that the SEP and non-SEP classes are almost
indistinguishable when all the dimensions are used. When the decision tree
feature selection is applied, the data points become more scattered in space and
therefore easier for the classifier to distinguish. We also note that both
decision tree classifiers have as a root a proton x-ray correlation parameter
($P6\_xl\_l2$). Some of the intermediate and leaf nodes have features that show
correlations between proton channels is their conditions. This suggests that
cross-correlations in proton channels are equally important to X-ray and proton
channels correlations that appeared in \cite{nunez2011predicting}. Our best
model shows a descent accuracy that is comparable (3\% better) to the UMASEP
system that uses the same catalog. We also made sure that our model is not
biased towards the missing data of the lower energy channels P6 and P7 of
GOES-12 by choosing the same number of samples of positive and negative class
that happened during the GOES-12 coverage period.

\section{Conclusion}
In this paper we designed a new model to predict $>$100 MeV SEP events based on
GOES satellite X-ray and proton data. This is the first effort that explores
not only the dependencies between the X-ray and proton channels but also the
auto-correlations and cross-correlations within the proton channels. We have
found that proton channel cross-correlations based on a lag time (prior point
in time) can be an important precursor as to whether an SEP event may happen or
not. In particular, we started finding patterns starting from lag 5 and our best
models shows both that the correlation between proton channel $P6$ and X-ray
channel $xl$ is an important precursor to SEP events.
Because of the missing data due to the failure of the P6 and P7 proton
channels onboard GOES-12 we made sure that our dataset uses the same number of
positive and negative examples coming from GOES-12. To our knowledge this is
the first study that explores proton cross-channels correlations in order to
predict SEP events. As a future extension of this study we are interested in
doing ternary classification by further splitting the SEP class into impulsive
and gradual. We are also interested in real-time SEP event predictions for
practical applications of this research.


\section*{Acknowledgment}
We thank all those involved with the GOES missions as well as the SOHO
mission.
We also acknowledge all the efforts of NOAA in making the catalogs and X-ray
reports available.
This work was supported in part by two NASA Grant Awards (No.
NNX11AM13A, and No. NNX15AF39G), and one NSF Grant (No. AC1443061). The
NSF Grant has been supported by funding from the Division of Advanced
Cyberinfrastructure within the Directorate for Computer and Information Science
and Engineering, the Division of Astronomical Sciences within the Directorate
for Mathematical and Physical Sciences, and the Division of Atmospheric and
Geospace Sciences within the Directorate for Geosciences.

\bibliographystyle{IEEEtran}
\bibliography{bib}

\begin{thebibliography}{10}
\providecommand{\url}[1]{#1}
\csname url@samestyle\endcsname
\providecommand{\newblock}{\relax}
\providecommand{\bibinfo}[2]{#2}
\providecommand{\BIBentrySTDinterwordspacing}{\spaceskip=0pt\relax}
\providecommand{\BIBentryALTinterwordstretchfactor}{4}
\providecommand{\BIBentryALTinterwordspacing}{\spaceskip=\fontdimen2\font plus
\BIBentryALTinterwordstretchfactor\fontdimen3\font minus
  \fontdimen4\font\relax}
\providecommand{\BIBforeignlanguage}[2]{{%
\expandafter\ifx\csname l@#1\endcsname\relax
\typeout{** WARNING: IEEEtran.bst: No hyphenation pattern has been}%
\typeout{** loaded for the language `#1'. Using the pattern for}%
\typeout{** the default language instead.}%
\else
\language=\csname l@#1\endcsname
\fi
#2}}
\providecommand{\BIBdecl}{\relax}
\BIBdecl

\bibitem{posner2007up}
A.~Posner, ``Up to 1-hour forecasting of radiation hazards from solar energetic
  ion events with relativistic electrons,'' \emph{Space Weather}, vol.~5,
  no.~5, 2007.

\bibitem{2016l}
M.~Desai and J.~Giacalone, ``Large gradual solar energetic particle events,''
  \emph{Living Reviews in Solar Physics}, vol.~13, no.~1, p.~3, 2016.

\bibitem{greendale}
\BIBentryALTinterwordspacing
``Space weather,'' accessed on 11-21-2017. [Online]. Available:
  \url{http://arcturan.com/space-weather/}
\BIBentrySTDinterwordspacing

\bibitem{gabriel1996power}
S.~Gabriel and J.~Feynman, ``Power-law distribution for solar energetic proton
  events,'' \emph{Solar Physics}, vol. 165, no.~2, pp. 337--346, 1996.

\bibitem{reames1999particle}
D.~V. Reames, ``Particle acceleration at the sun and in the heliosphere,''
  \emph{Space Science Reviews}, vol.~90, no. 3-4, pp. 413--491, 1999.

\bibitem{p0}
J.~Luhmann, S.~Ledvina, D.~Odstrcil, M.~J. Owens, X.-P. Zhao, Y.~Liu, and
  P.~Riley, ``Cone model-based {SEP} event calculations for applications to
  multipoint observations,'' \emph{Advances in Space Research}, vol.~46, no.~1,
  pp. 1--21, 2010.

\bibitem{p1}
J.~B. Robinson, ``Energy backcasting a proposed method of policy analysis,''
  \emph{Energy policy}, vol.~10, no.~4, pp. 337--344, 1982.

\bibitem{c0}
A.~Anastasiadis, A.~Papaioannou, I.~Sandberg, M.~Georgoulis, K.~Tziotziou,
  A.~Kouloumvakos, and P.~Jiggens, ``Predicting flares and solar energetic
  particle events: The {FORSPEF} tool,'' \emph{Solar Physics}, vol. 292, no.~9,
  p. 134, 2017.

\bibitem{c1}
M.~Dierckxsens, K.~Tziotziou, S.~Dalla, I.~Patsou, M.~Marsh, N.~Crosby,
  O.~Malandraki, and N.~Lygeros, ``The {COME{SEP} {SEP}} forecast tool,'' in
  \emph{EGU General Assembly Conference Abstracts}, vol.~16, 2014.

\bibitem{aran2006solpenco}
A.~Aran, B.~Sanahuja, and D.~Lario, ``Solpenco: A solar particle engineering
  code,'' \emph{Advances in Space Research}, vol.~37, no.~6, pp. 1240--1246,
  2006.

\bibitem{PPS}
S.~Kahler, E.~Cliver, and A.~Ling, ``Validating the proton prediction system
  {(PPS}),'' \emph{Journal of atmospheric and solar-terrestrial physics},
  vol.~69, no.~1, pp. 43--49, 2007.

\bibitem{laurenza2009technique}
M.~Laurenza, E.~Cliver, J.~Hewitt, M.~Storini, A.~Ling, C.~Balch, and
  M.~Kaiser, ``A technique for short-term warning of solar energetic particle
  events based on flare location, flare size, and evidence of particle
  escape,'' \emph{Space Weather}, vol.~7, no.~4, 2009.

\bibitem{neal2001predicting}
J.~S. Neal and L.~W. Townsend, ``Predicting dose-time profiles of solar
  energetic particle events using bayesian forecasting methods,'' \emph{IEEE
  transactions on nuclear science}, vol.~48, no.~6, pp. 2004--2009, 2001.

\bibitem{nunez2015real}
M.~N{\'u}{\~n}ez, ``Real-time prediction of the occurrence and intensity of the
  first hours of \textgreater 100 {M}e{V} solar energetic proton events,''
  \emph{Space Weather}, vol.~13, no.~11, pp. 807--819, 2015.

\bibitem{nunez2011predicting}
M.~Nunez, ``Predicting solar energetic proton events ({E} \textgreater 10
  {M}e{V}),'' \emph{Space Weather}, vol.~9, no.~7, 2011.

\bibitem{xi2006fast}
X.~Xi, E.~Keogh, C.~Shelton, L.~Wei, and C.~A. Ratanamahatana, ``Fast time
  series classification using numerosity reduction,'' in \emph{Proceedings of
  the 23rd international conference on Machine learning}.\hskip 1em plus 0.5em
  minus 0.4em\relax ACM, 2006, pp. 1033--1040.

\bibitem{scargle1982studies}
J.~D. Scargle, ``Studies in astronomical time series analysis. ii-statistical
  aspects of spectral analysis of unevenly spaced data,'' \emph{The
  Astrophysical Journal}, vol. 263, pp. 835--853, 1982.

\bibitem{zivot2006vector}
E.~Zivot and J.~Wang, ``Vector autoregressive models for multivariate time
  series,'' \emph{Modeling Financial Time Series with S-Plus{\textregistered}},
  pp. 385--429, 2006.

\bibitem{safavian1991survey}
S.~R. Safavian and D.~Landgrebe, ``A survey of decision tree classifier
  methodology,'' \emph{IEEE transactions on systems, man, and cybernetics},
  vol.~21, no.~3, pp. 660--674, 1991.

\bibitem{loh2011classification}
W.-Y. Loh, ``Classification and regression trees,'' \emph{Wiley
  Interdisciplinary Reviews: Data Mining and Knowledge Discovery}, vol.~1,
  no.~1, pp. 14--23, 2011.

\bibitem{steinberg2009cart}
D.~Steinberg and P.~Colla, ``{CART}: classification and regression trees,''
  \emph{The top ten algorithms in data mining}, vol.~9, p. 179, 2009.

\bibitem{madhavan1997new}
P.~Madhavan, ``A new recurrent neural network learning algorithm for time
  series prediction,'' \emph{Journal of Intelligent Systems}, vol.~7, no. 1-2,
  pp. 103--116, 1997.

\bibitem{pedregosa2011scikit}
F.~Pedregosa, G.~Varoquaux, A.~Gramfort, V.~Michel, B.~Thirion, O.~Grisel,
  M.~Blondel, P.~Prettenhofer, R.~Weiss, V.~Dubourg \emph{et~al.},
  ``{Scikit-learn}: Machine learning in python,'' \emph{Journal of Machine
  Learning Research}, vol.~12, no. Oct, pp. 2825--2830, 2011.

\bibitem{auccp}
C.~X. Ling, J.~Huang, and H.~Zhang, ``{AUC}: a statistically consistent and
  more discriminating measure than accuracy,'' in \emph{IJCAI}, vol.~3, 2003,
  pp. 519--524.

\bibitem{PCA}
S.~Wold, K.~Esbensen, and P.~Geladi, ``Principal component analysis,''
  \emph{Chemometrics and intelligent laboratory systems}, vol.~2, no. 1-3, pp.
  37--52, 1987.

\end{thebibliography}

\end{document}